\documentclass[a4paper]{article}

\usepackage[english]{babel}
\usepackage[utf8x]{inputenc}
\usepackage[T1]{fontenc}
\usepackage{slashed}

\usepackage[a4paper,top=3cm,bottom=2cm,left=3cm,right=3cm,marginparwidth=1.75cm]{geometry}

\usepackage{svg}
\usepackage{bbm}
\usepackage{wrapfig}
\usepackage{listings}
\usepackage{color}
\definecolor{mygreen}{RGB}{28,172,0}
\definecolor{mylilas}{RGB}{170,55,241}
\usepackage{graphicx}
\usepackage{amsfonts}
\usepackage{booktabs}
\usepackage{siunitx}
\usepackage{caption}

\usepackage{float}
\usepackage{blindtext}
\usepackage{amsmath}
\usepackage{subcaption}
\usepackage{graphicx}
\usepackage{wasysym}
\usepackage{url}
\usepackage[colorlinks=true, allcolors=blue]{hyperref}

\numberwithin{equation}{section}

\newcommand{\ist}[1]{\overset{\footnotesize(\ref{#1})}{=}}
\newcommand{\tist}[1]{\overset{\footnotesize #1}{=}}
\newcommand{\iist}[2]{\overset{^{(\ref{#1})}}{\underset{^{(\ref{#2})}}{=}}}
\newcommand{\zu}[1]{\stackrel{\footnotesize(\ref{#1})}{\mapsto}}
\newcommand{\gehtzu}[1]{\stackrel{\footnotesize#1}{\to}}

\title{\textbf{Numerical Evaluation of a Soliton Pair with Long Range Interaction}}
\author{\textbf{Joachim Wabnig, Josef Resch, Dominik Theuerkauf,}\\\textbf{Fabian Anmasser  and Manfried Faber}}

\begin{document}
\maketitle

\begin{abstract}
Within the model of topological particles (MTP) we determine the interaction energy of monopole pairs, sources and sinks of a Coulombic field. The monopoles are represented by topological solitons of finite size and mass, described by a field without any divergences. We fix the soliton centres in numerical calculations at varying distance. Due to the finite size of the solitons we get deviations from the Coulomb potential at distances of a few soliton radii. We compare the numerical results for these deviations with the running of the coupling in perturbative QED.
\end{abstract}

\section{Introduction}

In Refs.~\cite{Faber:1999ia,Faber:2022zwv} we proposed a dynamical model for monopoles without any singularities, the model of topological particles (MTP). Particles are identified by topological quantum numbers, masses of particles originate in field energy, charges show Coulombic behaviour and are quantised in units of an elementary charge. Depending on the interpretation, charges can be either electric or magnetic. The model and its many predictions were extensively discussed in Refs.~\cite{Faber:2022zwv}.

Magnetic monopoles were invented by Dirac in 1931~\cite{dirac:1931kp,dirac:1948um} as quantised singularities in the electromagnetic field. He found that their existence would explain the quantisation of electric charge, proven in Millikan's experiments~\cite{Millikan1913} but not explained by Maxwell theory~\cite{Jackson:1999cl}. Dirac monopoles have two types of singularities, the Dirac string, a line-like singularity connecting monopoles and antimonopoles, and the singularity in the centre of the monopole, a singularity analogous to the singularity of point-like electrons. Wu and Yang succeeded to formulate magnetic monopoles without the line-like singularities of the Dirac strings by using either a fibre-bundle construction with two different gauge fields~\cite{Wu:1975es}, one for the northern and one for the southern hemisphere of the monopole, or by a non-abelian SU(2) gauge field in 3+1D~\cite{Wu:1969wy,Wu:1975vq}. These Wu-Yang monopoles still suffer from the point-like singularities in the centre. There are monopole solutions without any singularity in the Georgi-Glashow model \cite{Georgi:1972cj}, the 'tHooft-Polyakov monopoles \cite{'tHooft:1974qc,Polyakov:1974ek}. The Georgi-Glashow model, formulated in 3+1D, has 15 degrees of freedom, an adjoint Higgs field with three degrees of freedom and an SU(2) gauge field with $4\cdot3=12$ field components. Only one degree of freedom is needed for the Sine-Gordon model \cite{Remoissenet:1999wa}, a model in 1+1D. It is most interesting that in addition to waves, it has kink and anti-kink solutions which interact with each other. The kink-antikink configurations are attracting and the kink-kink configurations repelling. The simplicity of the Sine-Gordon model inspired Skyrme \cite{Skyrme:1958vn,Skyrme:1961vq,Skyrme:1962vh,Makhankov:93sm} to a model in 3+1D with a scalar SU(2)-valued field, i.e. three degrees of freedom. Stable topological solitons (Skyrmions) emerge in that model with the properties of particles, interacting at short range. MTP was inspired by the simplicity and physical content of the Sine-Gordon model, it was first formulated in \cite{Faber:1999ia}. It has the same degrees of freedom as the Skyrme model, but uses a different Lagrangian. Its relations to electrodynamics and symmetry breaking were discussed in \cite{Faber:2022zwv,Faber:2002nw,Borisyuk:2007bd,Faber:2008hr}.

In this article, we want to concentrate on numerical determinations of the interaction energy for a pair of charges, represented by a soliton-antisoliton pair. Due to the non-existence of magnetic monopoles, comparisons to nature are possible only for electric charges and the predictions of QED.

In Sect.~\ref{Sec:Formulation} we repeat the formulation and some basic properties of the model, in Sect.~\ref{Sec:Lattice} we present the numerical formulation in cylindrical coordinates and in Sect.~\ref{Sec:Results} we show first results of the calculations and compare the results to perturbative QED.

\section{Summary of the MTP}\label{Sec:Formulation}
As described in Ref.~\cite{Faber:2022zwv} we use the SO(3) degrees of freedom of spatial Dreibeins to describe electromagnetic phenomena. The calculations get simpler using SU(2) matrices,
\begin{equation}\label{FeldVariablen}
Q(x)=\mathrm e^{-\mathrm i\alpha(x)\vec\sigma\vec n(x)}
=\underbrace{\cos\alpha(x)}_{q_0(x)}-\mathrm i\vec\sigma\underbrace{\vec n\sin\alpha(x)}_{\vec q(x)}\in SU(2)\cong\mathbb S^3
\end{equation}
in Minkowski space-time as field variables, where arrows indicate vectors in the 3D algebra of su(2) with the basis vectors represented by the Pauli matrices $\sigma_i$. The Lagrangian of MTP reads,
\begin{equation}
\label{LagrangianMTP}
\mathcal L_{\mathrm{MTP}}(x):=-\frac{\alpha_f\hbar c}{4\pi}
\left(\frac{1}{4}\,\vec R_{\mu\nu}(x)\vec R^{\mu\nu}(x)+\Lambda(x)\right)
\quad\textrm{with}\quad\Lambda(x):=\frac{q_0^6(x)}{r_0^4},
\end{equation}
with
\begin{equation}\label{curvaturetensor}
\vec R_{\mu\nu}:=\vec{\Gamma}_\mu\times\vec{\Gamma}_\nu\quad\textrm{and}\quad
\left(\partial_\mu Q\right)Q^\dagger=:-\mathrm i\vec\sigma\vec\Gamma_\mu.
\end{equation}

We get contact with nature by relating the electromagnetic field strength tensor to the dual of the curvature tensor $\vec R_{\mu\nu}$,
\begin{equation}\label{FieldStrength}
\vec F_{\mu\nu}:=-\frac{e_0}{4\pi\epsilon_0}^{\star\!\!}\vec R_{\mu\nu}.
\end{equation}
At large distances $d$ from the sources, measured in units of the soliton scale parameter $r_0$, the non-abelian field strength gets abelian,
\begin{equation}\label{asymFeld}
\vec F_{\mu\nu}\gehtzu{d>>r_0}(\vec F_{\mu\nu}\vec n)\vec n.
\end{equation}

MTP has four different classes of topologically stable single soliton configurations. Their representatives read,
\begin{equation}\label{Igel}
n_i(x)=\pm\frac{x^i}{r},\quad\alpha(x)=\frac{\pi}{2}\mp\arctan\frac{r_0}{r}
=\begin{cases}\arctan\frac{r}{r_0}\\\pi-\arctan\frac{r}{r_0}\end{cases}
\end{equation}
The imaginary part of their Q-fields are schematically depicted in Table~\ref{TabSign}.
\begin{table}[h]\begin{center}\hspace*{-4.5mm}\begin{tabular}{cccc}\hline 
$\vec n=\vec r/r$&$\vec n=-\vec r/r$&-$\vec n=\vec r/r$&$\vec n=\vec r/r$\\
$q_0\ge 0$&$q_0\le 0$&$q_0\ge 0$&$q_0\le 0$\\\hline
$Z=1$&$Z=-1$&$Z=-1$&$Z=1$\\
$\mathcal Q=\frac{1}{2}$&$\mathcal Q=\frac{1}{2}$&$\mathcal Q=-\frac{1}{2}$&$\mathcal Q=-\frac{1}{2}$\\\hline
\includegraphics[scale=0.18]{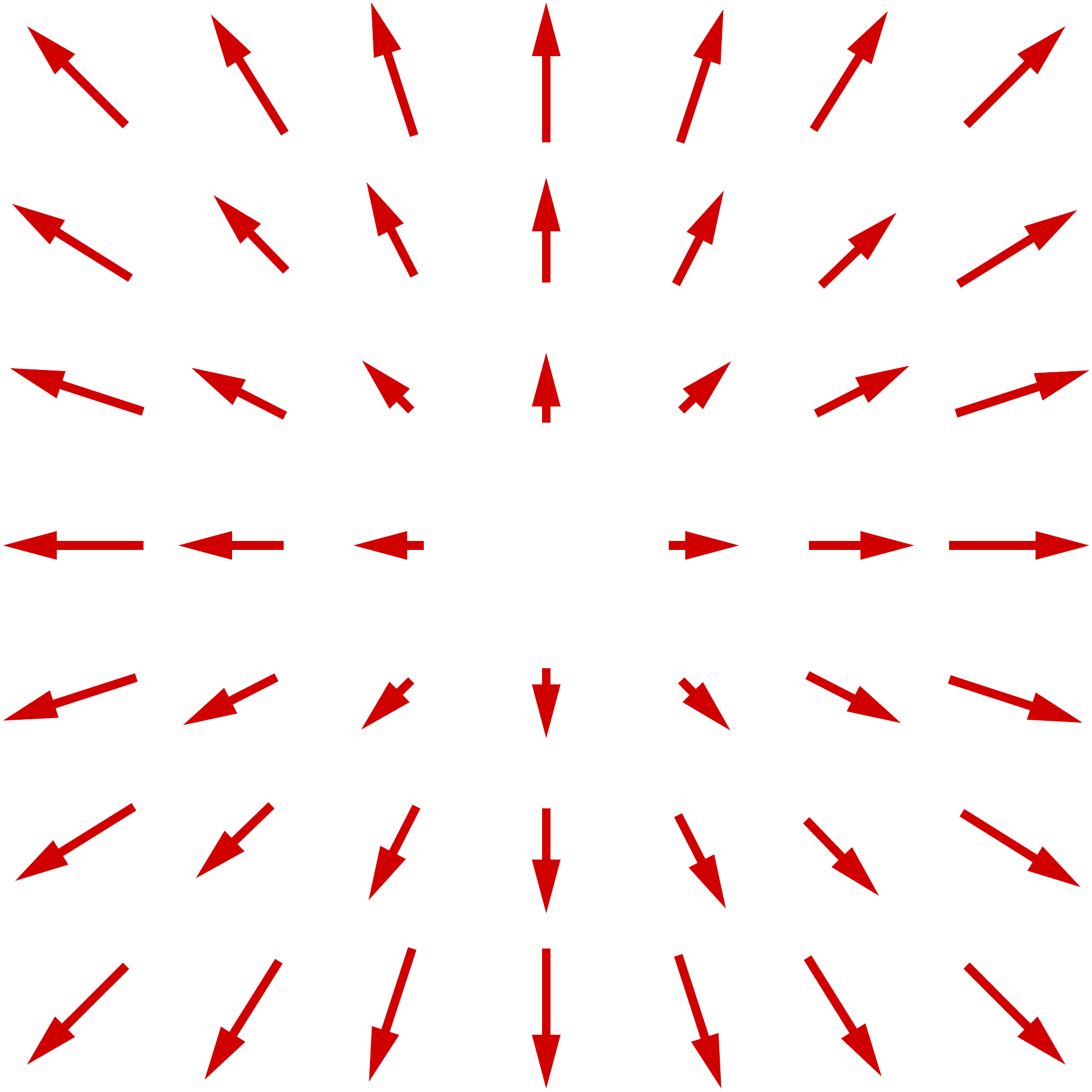}&\includegraphics[scale=0.18]{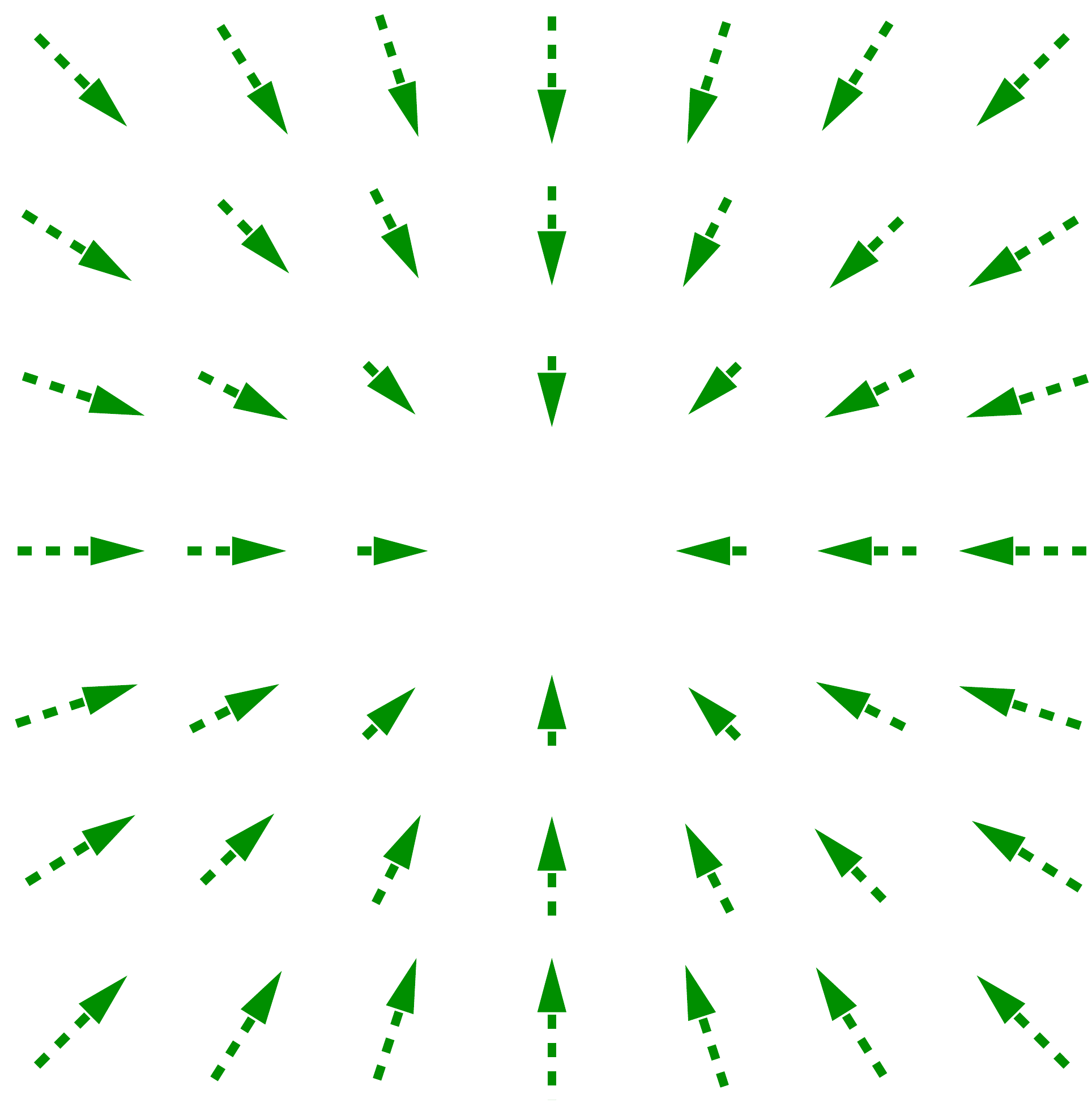}&\includegraphics[scale=0.18]{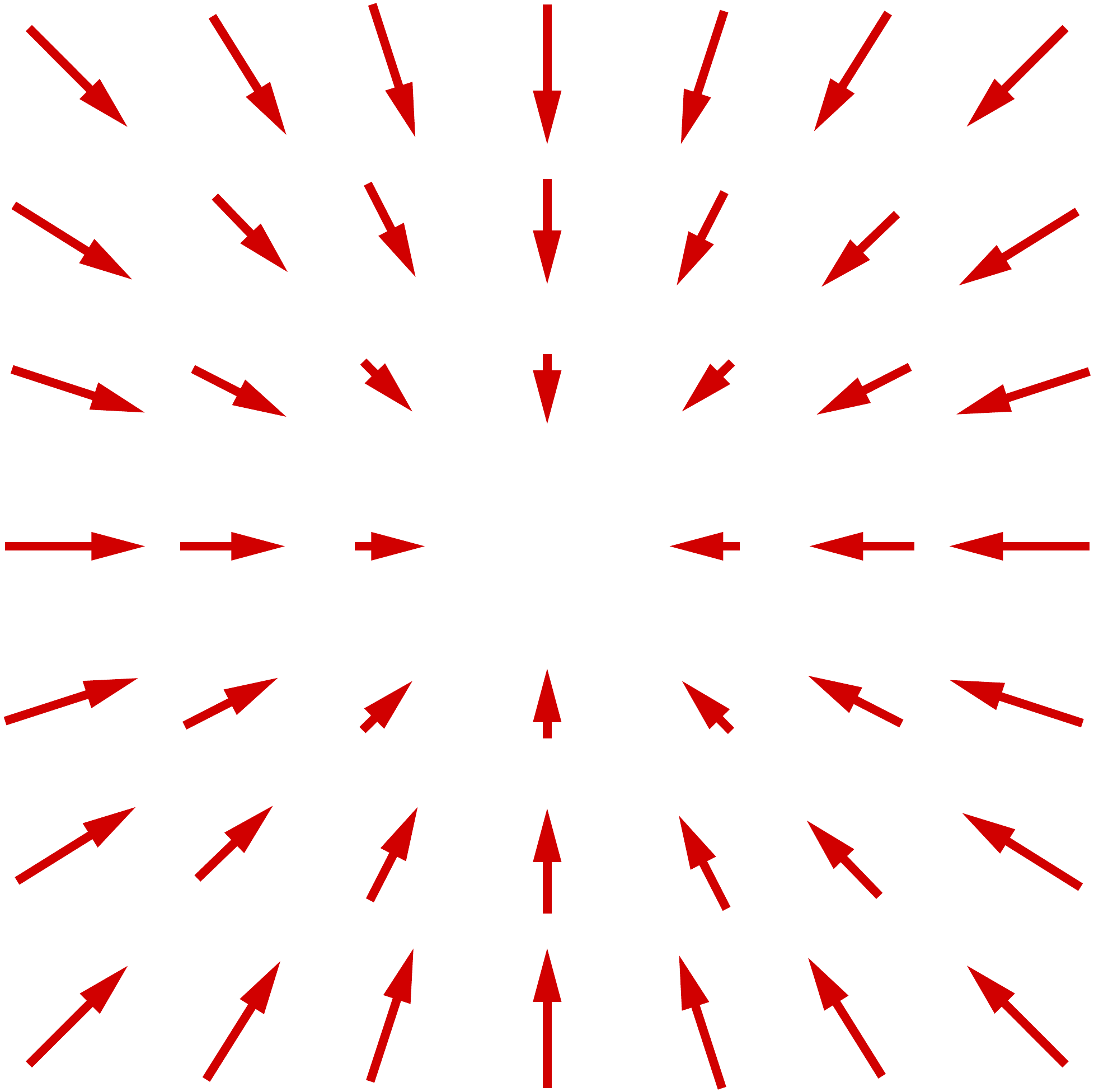}&\includegraphics[scale=0.18]{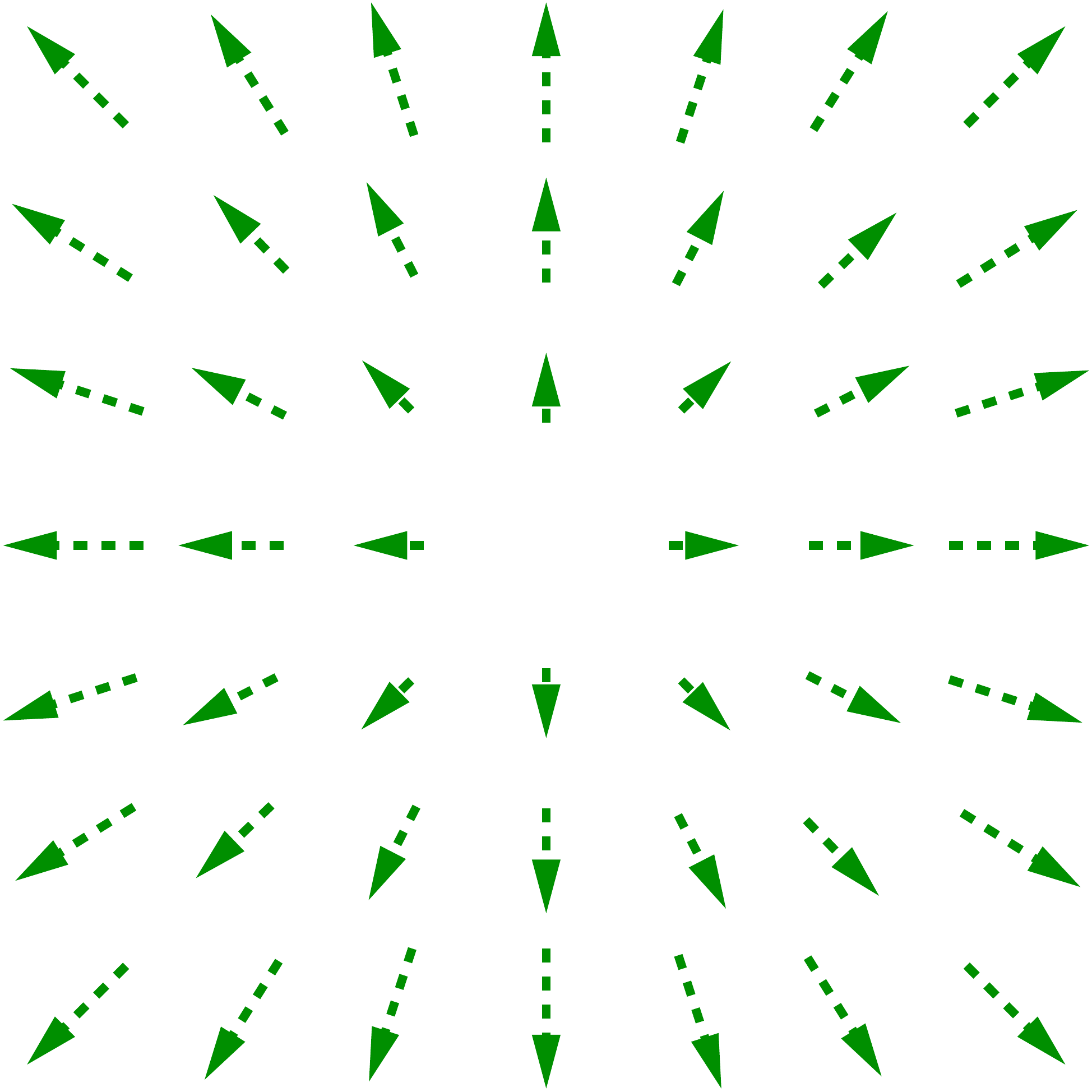}\\\hline
\end{tabular}\caption{Scheme of single soliton configurations. The fields $\vec n$ and $q_0$ and the topological quantum numbers $Z$ and $\mathcal Q$ are indicated. The diagrams show the imaginary components $\vec q=\vec n\sin\alpha$ of the soliton field, in full red for the hemisphere with $q_0>0$ and in dashed green for $q_0<0$.}
\label{TabSign}\end{center}\end{table}
The fields~(\ref{Igel}) are solutions of the non-linear Euler-Lagrange equations~\cite{Faber:2022zwv}. They differ in two quantum numbers related to charge and spin. In the minimum of the potential the $Q$-field is purely imaginary. Therefore, the sign $Z$ of the $\vec n$-field in Eq.~(\ref{Igel}) decides about the charge quantum number defined by a map $\Pi_2(\mathbb S^2)$. Field configurations are further characterised by the number $\mathcal Q$ of coverings of $\mathbb S^3$, by the map $\Pi_3(\mathbb S^3)$. With the sign of $\mathcal Q$ we define an internal chirality $\chi$ and with the absolute value of $\mathcal Q$ the spin quantum number $s$,
\begin{equation}\label{DefChiral}
\mathcal Q=\chi\cdot s\quad\textrm{with}\quad s:=|\mathcal Q|.
\end{equation}
The spin quantum number of two-soliton configurations indicates that $\chi$ can be related to the sign of the magnetic spin quantum number.

The configurations within each of the four classes may differ by Poincaré transformations. The rest mass of solitons,
\begin{equation}\label{eRuhenergie}
E_0=\frac{\alpha_f\hbar c}{r_0}\frac{\pi}{4},
\end{equation}
can be adjusted to the electron rest energy $m_ec_0^2=0.510\,998\,95$~MeV by choosing,
\begin{equation}\label{r0m2}
r_0=2.213\,205\,16~\mathrm{fm},
\end{equation}
a scale which is of the order of the classical electron radius. The four parameters $r_0, c_0, E_0$ and $e_0$ correspond to the natural scales of the four quantities length, time, mass and charge of the SI, of the Système international d’unités, involved in this model. Eq.~(\ref{eRuhenergie}) can therefore be interpreted as a relation between $\alpha_f$ and $\hbar$.

\begin{figure}[h!]
\hspace*{-5mm}\includegraphics[scale=0.6]{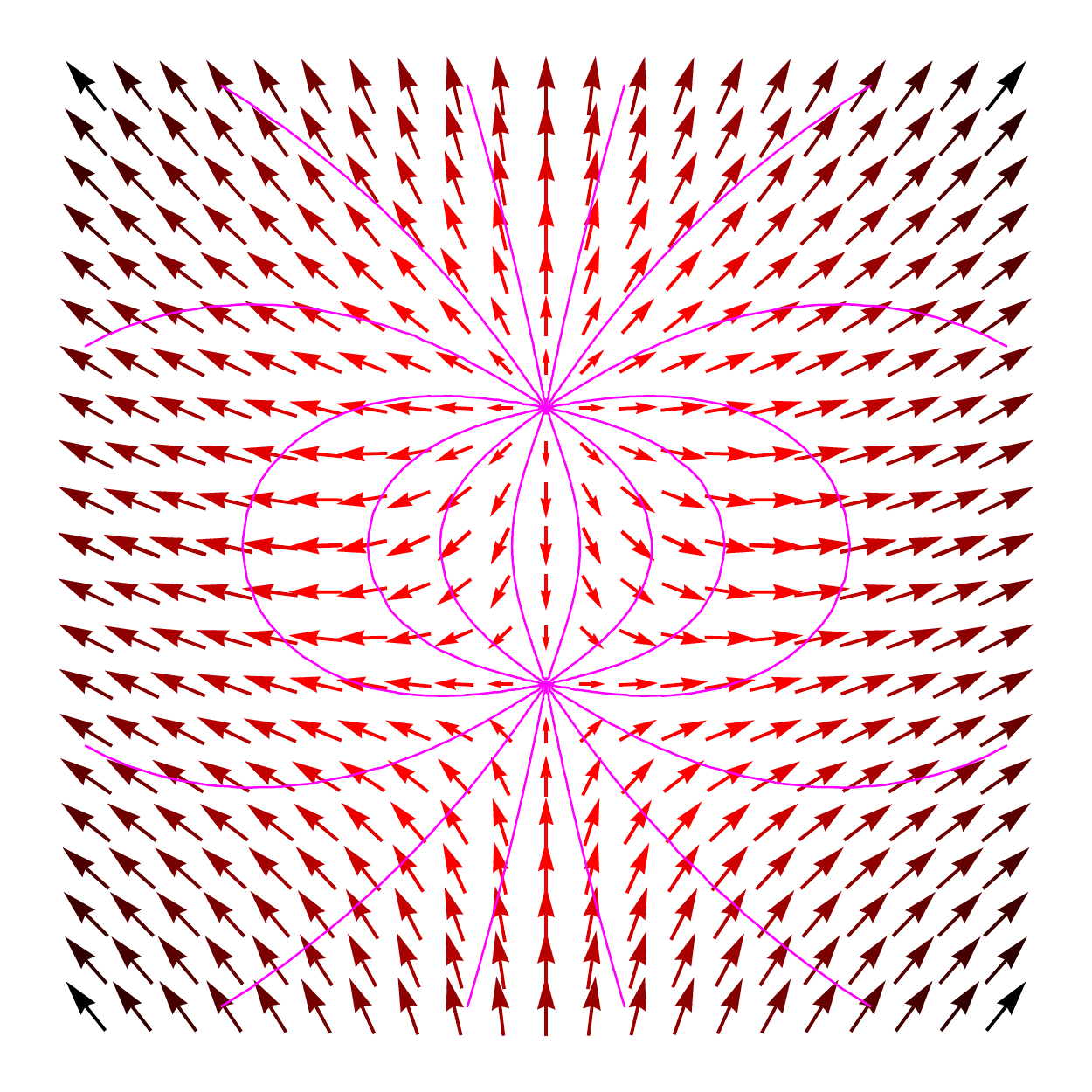}\hspace{5mm}
\includegraphics[scale=0.6]{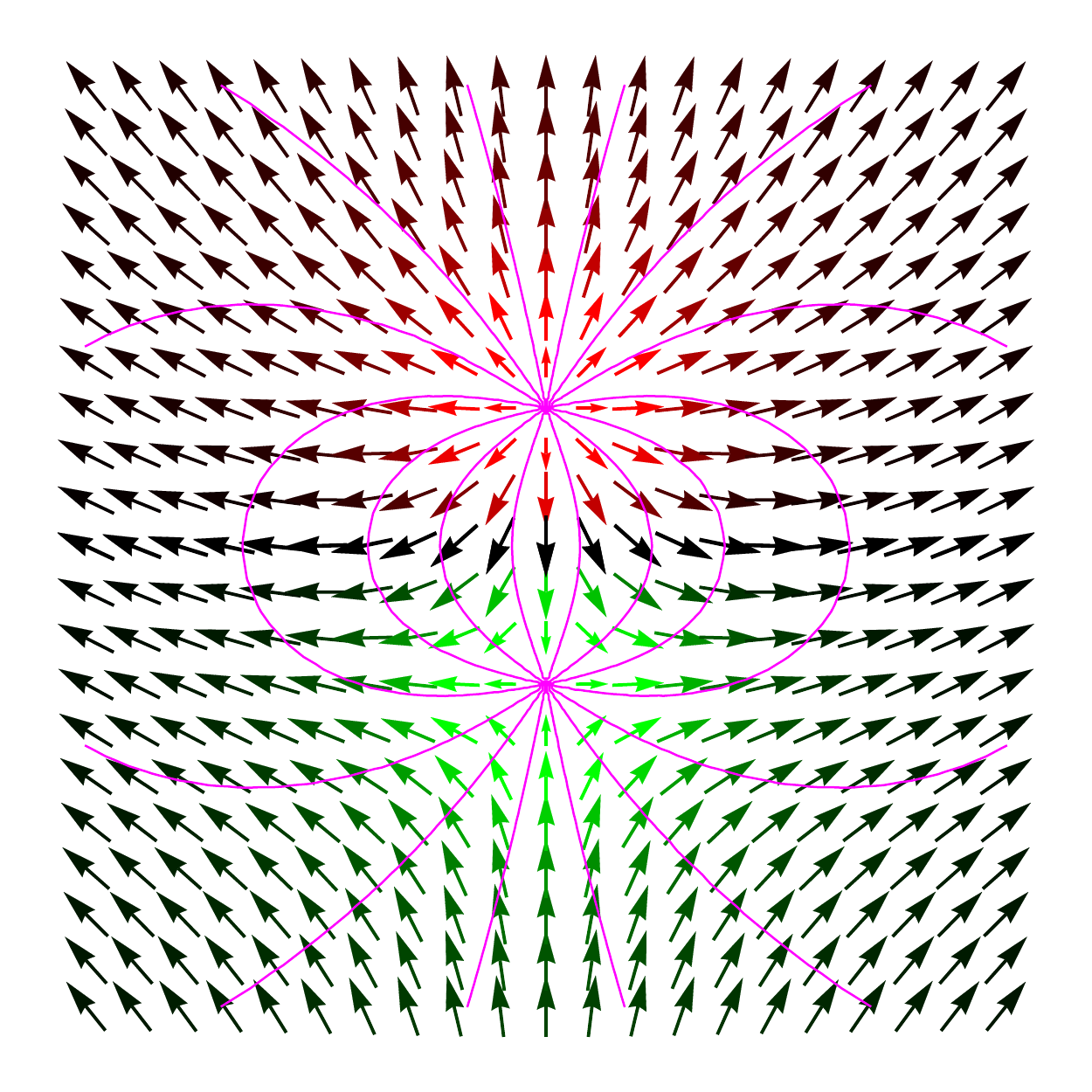}
\caption{Schematic diagrams depicting the imaginary part $\vec q=\vec n\sin\alpha$ of the $Q$-field of two opposite unit charges by arrows. The lines represent some electric flux lines. We observe that they coincide with the lines of constant $\vec n$-field. The configurations are rotational symmetric around the axis through the two charge centres. In the red/green arrows, we encrypt also the positive/negative values of $q_0=\cos\alpha$. For $q_0\to0$ the arrows are getting darker or black. The left configuration belongs to the topological quantum numbers $\mathcal Q=S=0$ and the right one to $\mathcal Q=S=1$, where $S$ is the total spin quantum number of the configuration. The numerical calculations will be presented for the spin singlet case.}
\label{anziehung}
\end{figure}

Two solitons of different charge can be combined to two topological different field configurations. Schematic diagrams for such configurations are shown in Fig.~\ref{anziehung}.

\section{Dipoles on a cylindrical lattice}\label{Sec:Lattice}
According to the Lagrangian~\ref{LagrangianMTP} there are two contributions to the energy density of a static dipole,
\begin{equation}\label{equ:EneDens}
\mathcal H\ist{LagrangianMTP}\frac{\alpha_f\hbar c}{4\pi}
\left(\frac{1}{4}\,\vec R_{ij}\vec R^{ij}+\Lambda\right)
=:\mathcal H_\mathrm{cur}+\mathcal H_\mathrm{pot},
\end{equation}
the electric part of the curvature energy and the potential energy. A detailed derivation of these energy contributions of the $Q(x)$ field in cylindrical coordinates was given in \cite{Anmasser2021}. The curvature energy reads in these coordinates,
\begin{equation}\begin{aligned}\label{curvatureEnergy}
\mathcal H_\mathrm{cur} = 
\frac{\epsilon_0}{2}&\big(\vec E_r^2+\vec E_\varphi^2+\vec E_z^2\big)
=\frac{\alpha_f\hbar c}{8\pi}\frac{1}{r^2}
\bigg\{q_r^2\big[(\partial_zq_0)^2+(\partial_zq_r)^2+(\partial_zq_z)^2\big]\\
&+\frac{r^2}{q_0^2}\big(\partial_rq_r\partial_zq_z
-\partial_zq_r\partial_rq_z\big)^2+q_r^2\big[(\partial_rq_0)^2
+(\partial_rq_r)^2+(\partial_rq_z)^2\big]\bigg\}.
\end{aligned}\end{equation}

\begin{figure}[htbp]
\centering
\includegraphics[width=0.75\textwidth]{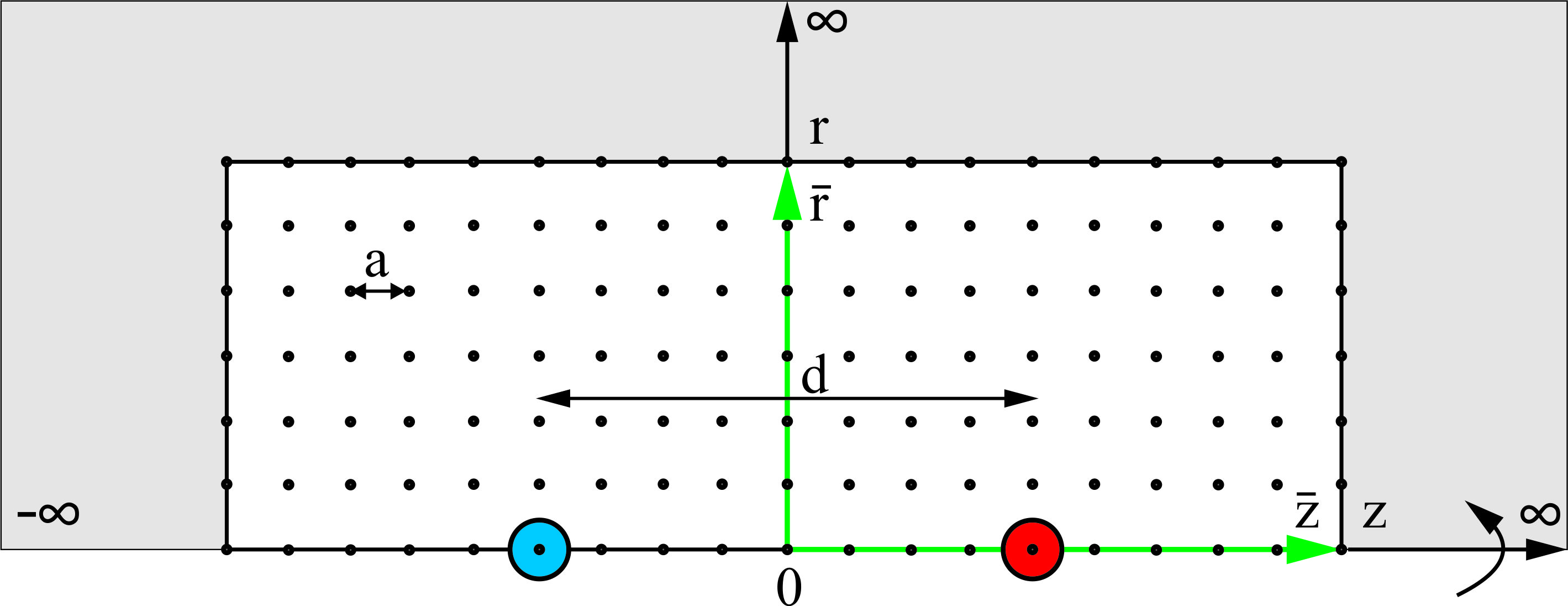}
\caption{Two solitons of opposite charge are separated by a distance $d$ on a discrete lattice in the $rz$ plane. The total number of points is $n_r \times 2n_z$ and the lattice constant in both directions is $a$.}
\label{dipoleScetch}
\end{figure}

Due to the cylindrical symmetry of the field configurations in Fig.~\ref{anziehung} it is sufficient to restrict the numerical integrations to a lattice in the $rz$-plane, see Fig.~\ref{dipoleScetch}.
Inside the lattice, we minimise the energy according to the MTP. Outside, we use Maxwell's electrodynamics~\cite{Jackson:1999cl} to calculate the energy. The numerical computations of dipole field configurations were built up on the algorithms discussed in \cite{Wabnig, Theuerkauf, Anmasser} and their accuracy was tested in \cite{Anmasser2021} by applying it to the analytical solvable monopole configuration. The coordinates in radial- and $z$ direction are denoted by $\bar{r}\in\{0,1,\dots,n_r\}$ and $\bar{z}\in\{-n_z,-n_z+1,\dots,\,n_z\}$, respectively. The radius of a soliton, $\bar{r}_0 \in \mathbb{N}$, determines the length-scale and thereby the lattice spacing,
\begin{equation}\label{Gitterkonst}
a\ist{r0m2}\frac{2.21\mathrm{fm}}{\bar{r}_0}.
\end{equation}
A numerical gradient minimization algorithm is used to find the dipole configurations, which minimizes the total energy according to Eq.~(\ref{equ:EneDens}) for given distance $d=a\bar d$ between the opposite charges.

A good guess for the initial configuration helps to reduce the run-time of the computation. For the real part of the initial $Q$-field we get from the profile function of a single monopole,
\begin{equation}\label{initq0}
q_0(r,z>0)=\frac{\bar{r}_0}{\sqrt{\bar{r}_0^2+\bar r^2+(\bar z-\frac{\bar d}{2})^2}}, \qquad
q_0(r,z \leq 0)=\frac{\bar{r}_0}{\sqrt{\bar{r}_0^2+\bar r^2+(\bar z+\frac{\bar d}{2})^2}}.
\end{equation}
The modulus of the vector part of the $Q$-field reads,
\begin{equation}
|\vec{q}(x)|\ist{FeldVariablen}\sqrt{1-q_0(x)^2}.
\end{equation}
An appropriate initial direction $\vec n$ of the $\vec{q}$ field we get from the coincidence between electric field lines and lines of constant $\vec n$-field which we observe in Fig.~\ref{anziehung}. The field lines are defined by the differential equation,
\begin{equation}\label{eqn:differ}
\frac{\mathrm dr}{E_r} = \frac{\mathrm dz}{E_z},
\end{equation}
with
\begin{equation}\begin{aligned}\label{Efeld}
E_{r}(r,z)&=\frac{e_0}{4\pi\epsilon_0} \left(\frac{r}{(r^2+(z+\frac{d}{2})^2)^{3/2}}-\frac{r}{(r^2+(z-\frac{d}{2})^2)^{3/2}}\right),\\
E_{z}(r, )&=\frac{e_0}{4\pi\epsilon_0} \left(\frac{z+\frac{d}{2}}{(r^2+(z+\frac{d}{2})^2)^{3/2}}-\frac{z-\frac{d}{2}}{(r^2+(z-\frac{d}{2})^2)^{3/2}}\right).
\end{aligned}\end{equation}
After solving the differential Eq. (\ref{eqn:differ}), we find an analytical expression for the initial polar angle of the $\vec{q}$ field,
\begin{equation}
\theta(r,z)=\arccos\left(1+\frac{z-\frac{d}{2}}{\sqrt{r^2+(z-\frac{d}{2})^2}}
-\frac{z+\frac{d}{2}}{\sqrt{r^2+(z+\frac{d}{2})^2}}\right).
\end{equation}

Up to now, only points on the lattice are included in the energy calculation. Outside the lattice, we use Maxwell's electrodynamics to take the curvature energy into account. The error, for neglecting potential and tangential energy components outside the lattice, was shown to be tiny ($<$ 0.2\%) for a single monopole, if the distance from the soliton core to the lattice boundaries is greater than $10a$ \cite{Anmasser2021}. Hence, we find the energy outside the lattice by $H_{\mathrm{out}} = \frac{\epsilon_0}{2} \int_{\mathbb{R}^3 \setminus \text{latt} } \mathrm{d}^3x |\vec{E}|^2$, which is numerically integrated using a trapezoidal summation, with the electric field strength given by Eq.~\ref{Efeld}.

The principle of energy minimisation is exploited to find the dipole field configuration and its associated energy. The minimisation algorithm is taken from \url{https://de.mathworks.com/matlabcentral/fileexchange/75546-conjugate-gradient-minimisation}. It is a more dimensional conjugate gradient method to find a local minimum of an energy functional depending on the $Q$ field components at each lattice point, $E=E(q_{0}^i, q_{r}^i, q_{z}^i)$ with $i \in [1,2,...,n_r(2n_z+1)]$,
\begin{equation}\label{Enumerisch}
E=\int_\mathrm{lattice}\mathrm d^3x~\mathcal H_\mathrm{lat}+H_\mathrm{out},
\end{equation}
where $\mathcal H_\mathrm{lat}$ is the lattice version of $\mathcal H$ of Eq.~(\ref{equ:EneDens}).

Fixing the centre of solitons only, however, for small dipole distances $d$ annihilation of the two solitons is observed. To suppress this behaviour, it was suggested in Ref.~\cite{Anmasser2021} to smooth the soliton field by adding to the energy functional a term,
\begin{equation}\label{massTerm}
E_\mathrm{min}:= E+\lambda\frac{\alpha_f\hbar c}{a}\sum_{i}
\left(\frac{\partial}{\partial\bar x_i}\frac{\vec q}{|\vec q|}\right)^2, \quad
i=\{\bar{r},\bar{z}\}.
\end{equation}
in the minimisation process, where $\lambda=\bar\lambda/a^2$. During the computations, it became clear that this additional term tends to suppress the interaction between the solitons. Therefore, we could apply tiny values of $\bar\lambda$ only. This was sufficient to approach smaller distances.

The parameters, used for the minimisation process are listed in table~\ref{Tafel}.
\begin{table}[h]
\begin{tabular}{ |l|c|r| } 
\hline
Parameter & Description & Used value\\
\hline
$i_{\mathrm{max}}$&Max. number of iterations & 5000\\
$\Delta_{\mathrm{grad}}$&Min. gradient difference for two consecutive iterations&1e-6\\
$\Delta_{\mathrm{fun}}$&Min. energy difference for two consecutive iterations&1e-30\\
$\Delta_{\mathrm{com}}$&Lower bound on the step size of the norm of the Q field & 1e-30
\\\hline
\end{tabular}
\caption{Input parameters for the energy minimisation algorithm.}
\label{Tafel}
\end{table}

The lattice spacing we fix to $a\ist{Gitterkonst}r_0/3=0.738$~fm. This spacing is small enough to get a reasonable approximation to the rest energy of electrons.

For a dipole, it is not possible to use periodic boundary conditions. While minimising the energy, we fix the soliton field $Q$ at the edge of the lattice $\bar{r}=0, \bar{r}=n_r, \bar{z}=\pm n_z$ according to the initial configuration discussed in Sect. \ref{Sec:Lattice}. This implies the assumption $q_0=0$ outside the lattice, and leads to a small error in the potential energy which is difficult to avoid.

Nevertheless, boundary conditions have large influence on the numerical results. For fixed lattice sizes $n_r\times (2n_z+1)$ and varying size $d$ of the dipole, we can not reach large $d$ values due to boundary effects. It turns out to be much better to increase the lattice size with increasing $d$ keeping the distance from the boundary constant. The lattice sizes we are using for our calculations are in r- and z-directions by 45 lattice spacings larger than the relative distance $d$ between the charges. We use $n_r=45$ and $n_z=\frac{\bar d}{2}+45$ and all even distances $\bar d\le420$.

Due to the finite size of the lattice there remains a deviation of the asymptotic energy $E_\infty$ of the dipole from the mass of two non-interacting solitons, $2m_ec_0^2=1.021\,997\,90$~MeV. Since we fix the centers of solitons at lattice sites, we get a weak even-odd effect on lattice sizes for $E_\infty$. For the asymptotic values $E_\infty$ of the calculated energies we get $1.009\,203$~MeV for even and $1.009\,198 $~MeV for odd $r$.

\begin{figure}[htbp]
\centering
\hspace*{-10mm}\includegraphics[scale=0.25]{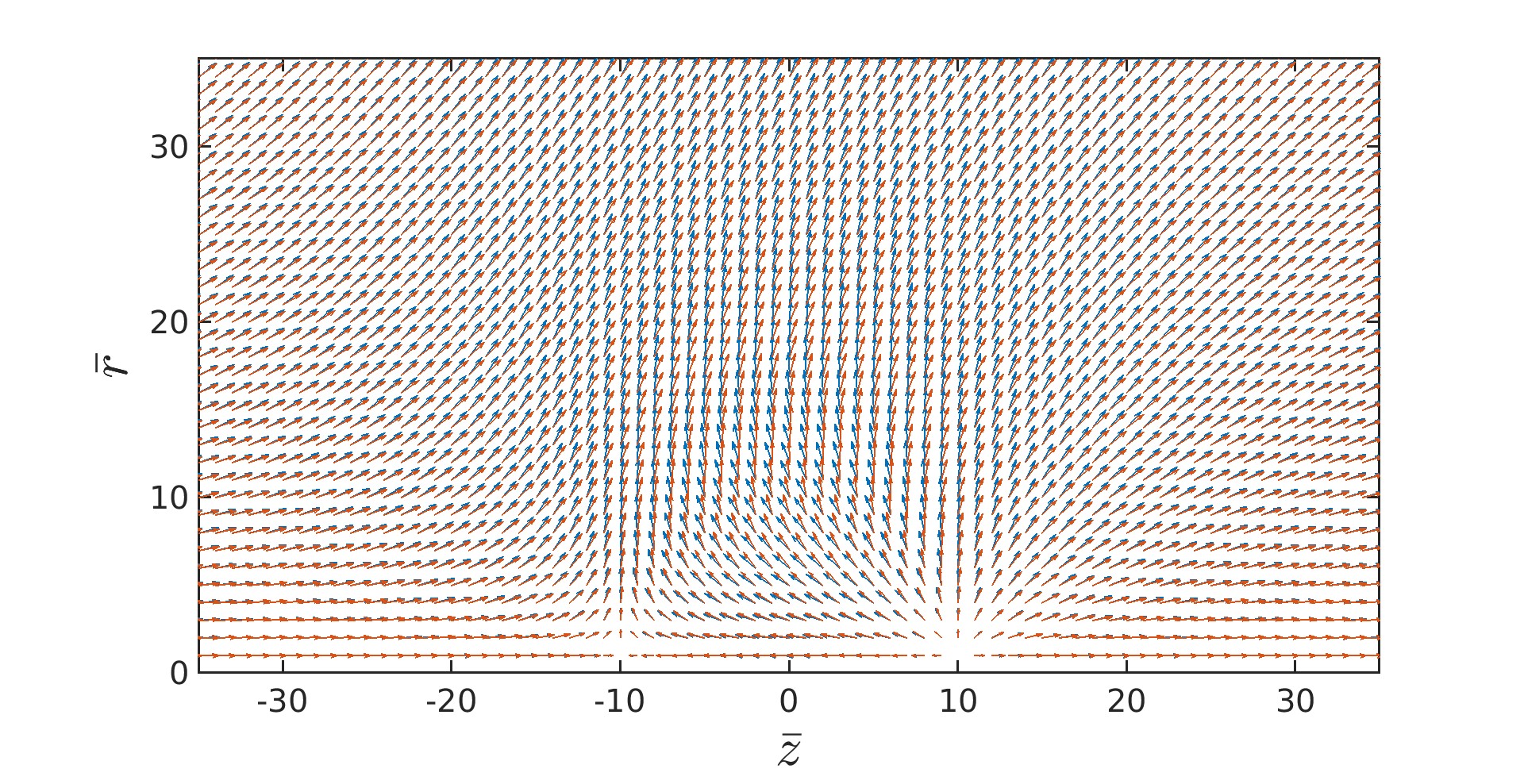}
\caption{The modification of the imaginary part $\vec{q}(r,z)$ of the soliton field is represented by arrows. The initial configuration is shown in blue and the result of the minimisation process in red. Only a subregion of the $45\times111$-lattice is shown. Due to the finite size of solitons, $\bar r_0=3$, the centres of the charges the length of the arrows approaches zero as described by Eq.~(\ref{Igel}).}
\label{fig:VecPartComp}
\end{figure}

\begin{figure}[htbp]
\centering
\includegraphics[width=1\textwidth]{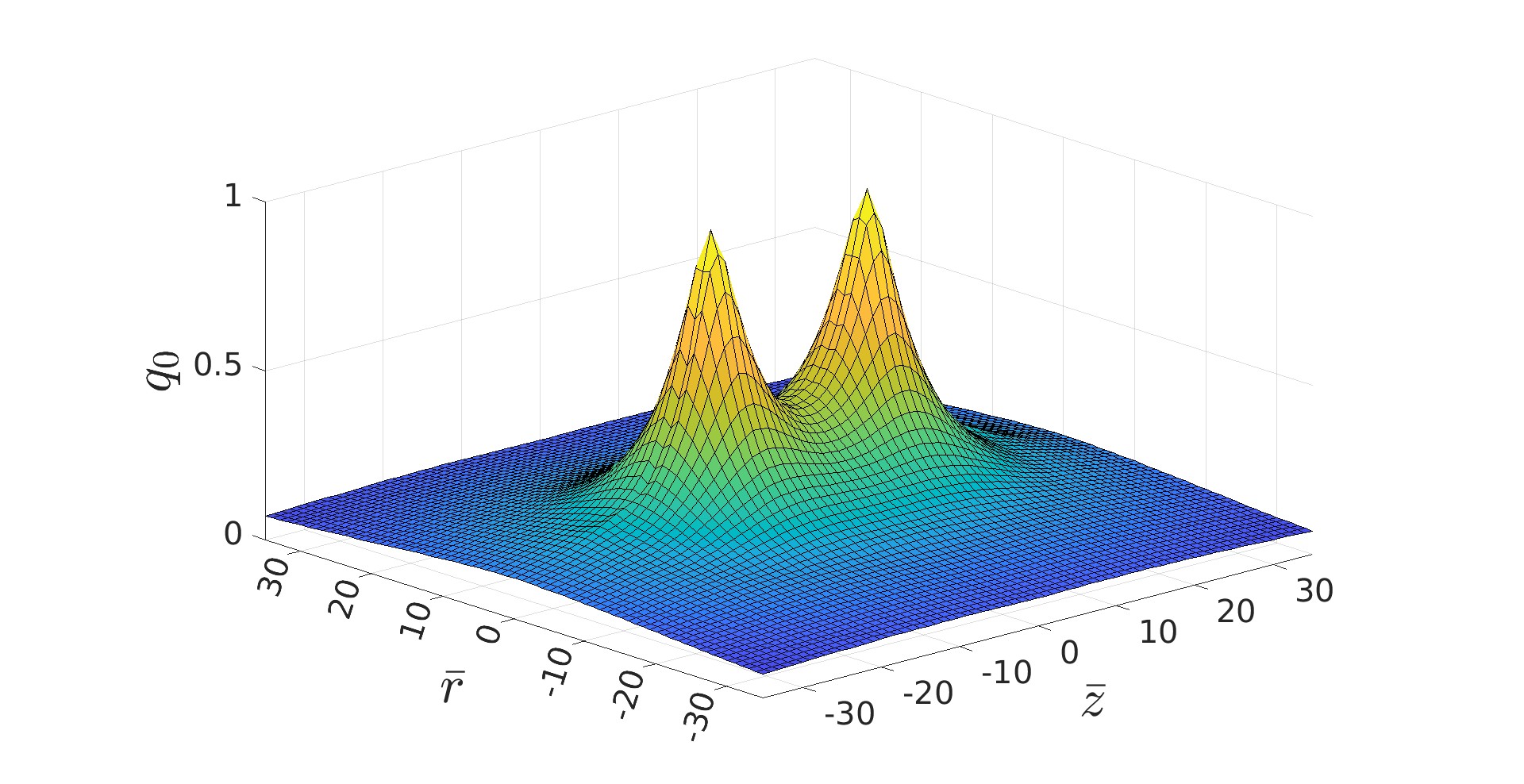}
\caption{$q_0$ distribution after minimisation for the configuration of Fig.~\ref{fig:VecPartComp}.}
\label{fig:q0min}
\end{figure}

\begin{figure}[htbp]
\centering
\includegraphics[width=1\textwidth]{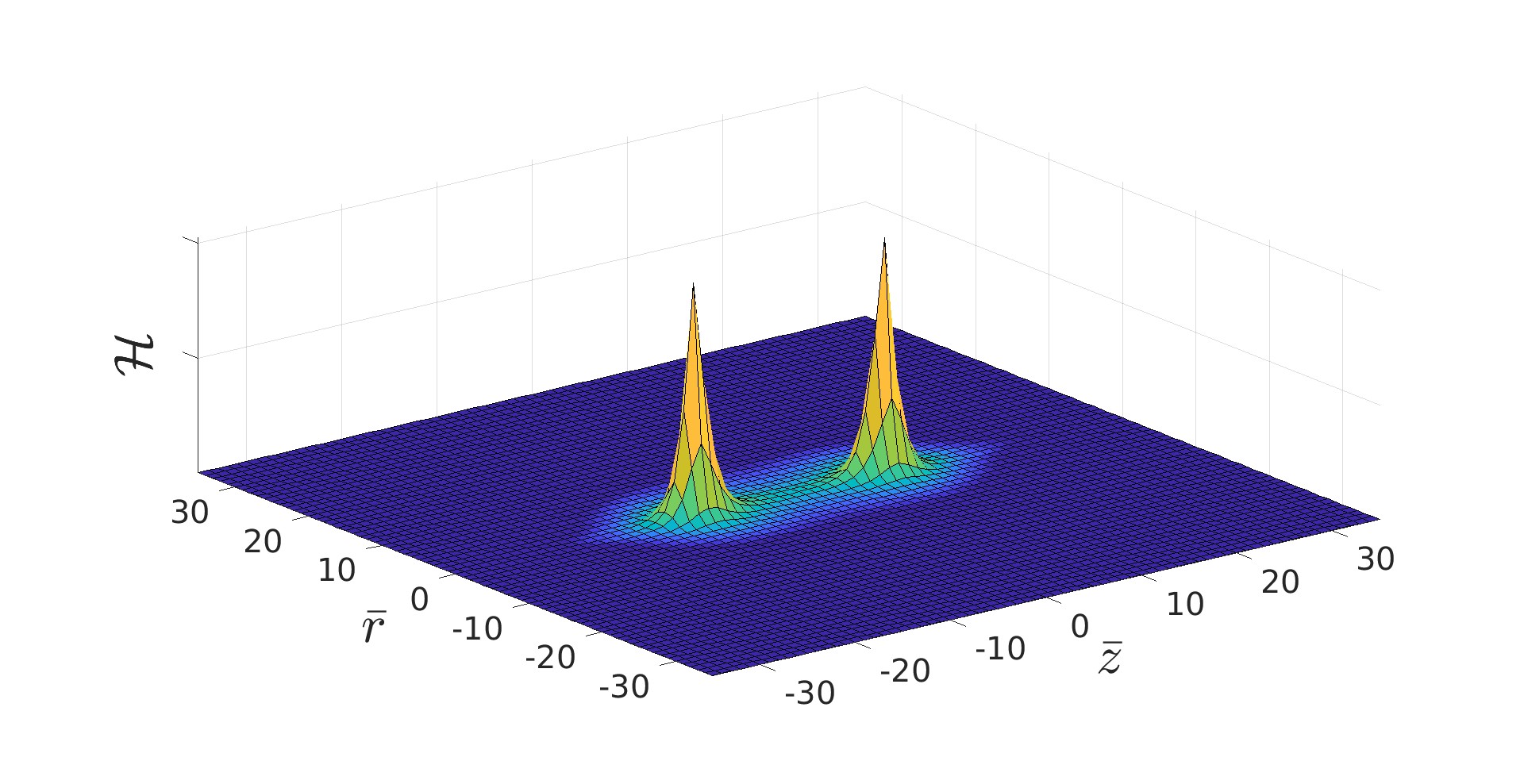}
\caption{Energy density after minimization for the configuration of Fig.~\ref{fig:VecPartComp}. Despite the regularisation of the self-energy of charges by the finite soliton size $\bar r_0=3$ there exist still very pronounced peaks of the energy density $\mathcal H$ according to Eq.~(\ref{equ:EneDens}) at the charge centres.}
\label{fig:Edens}
\end{figure}

For the distance $\bar d=16$ we show some results of the minimisation process for the spin singlet configuration. In Fig.~\ref{fig:VecPartComp} we compare the $\vec q$-field in the initial and final configuration in a restricted region.
The minimisation modifies especially the field in the region between the charges. The final values of the real part $q_0$ of the soliton field are shown in Fig.~\ref{fig:q0min}. We observe that the distribution of $q_0$ values seems much broader than the distribution of the energy density in Fig.~\ref{fig:Edens} with very pronounced peaks reflecting the self energy $\mathcal H$ of charges according to Eq.~(\ref{equ:EneDens}).

\section{Comparison to and deviations from Coulomb potential}\label{Sec:Results}
\begin{figure}[h]
\centering
\includegraphics[scale=0.25]{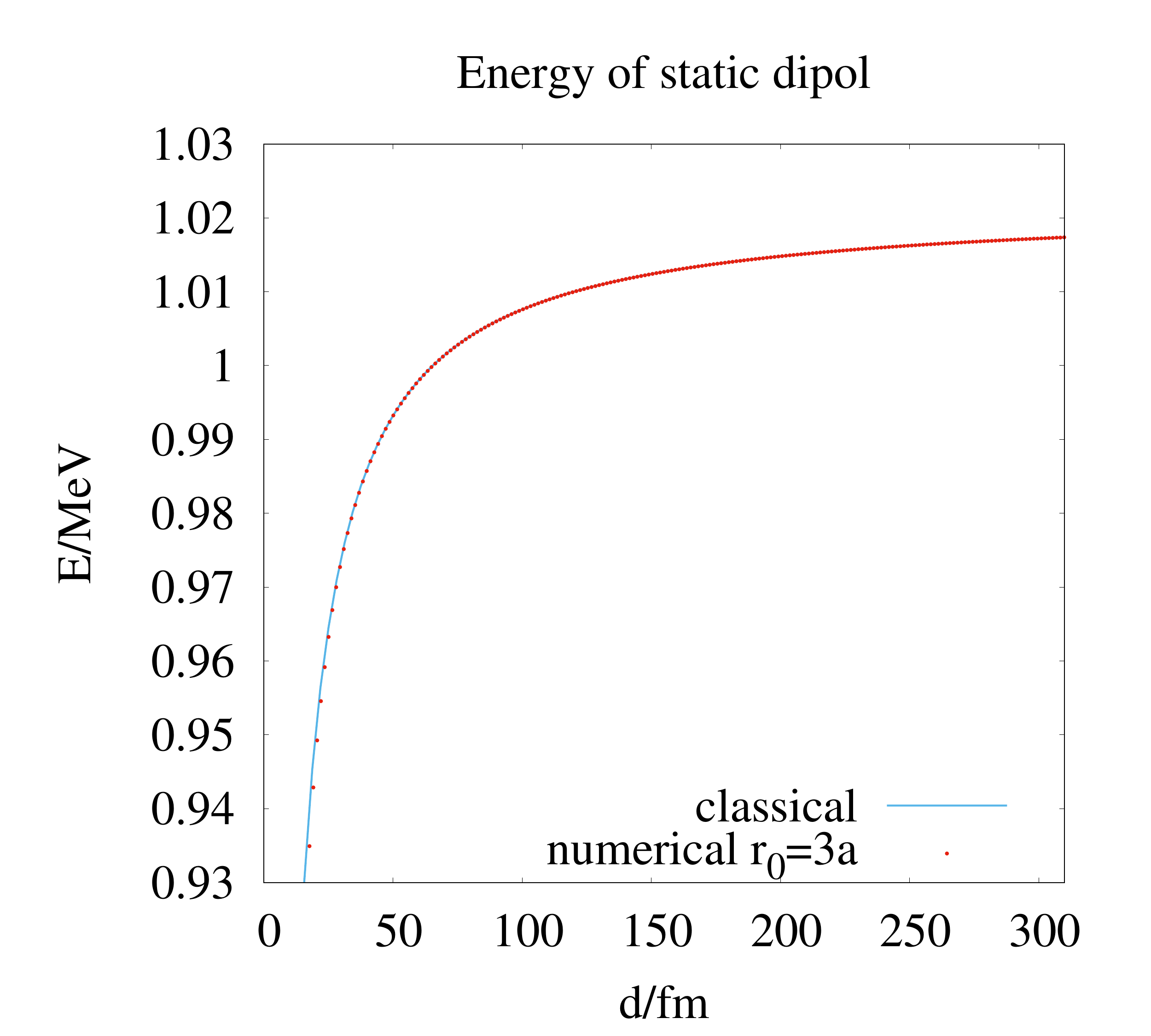}
\includegraphics[scale=0.25]{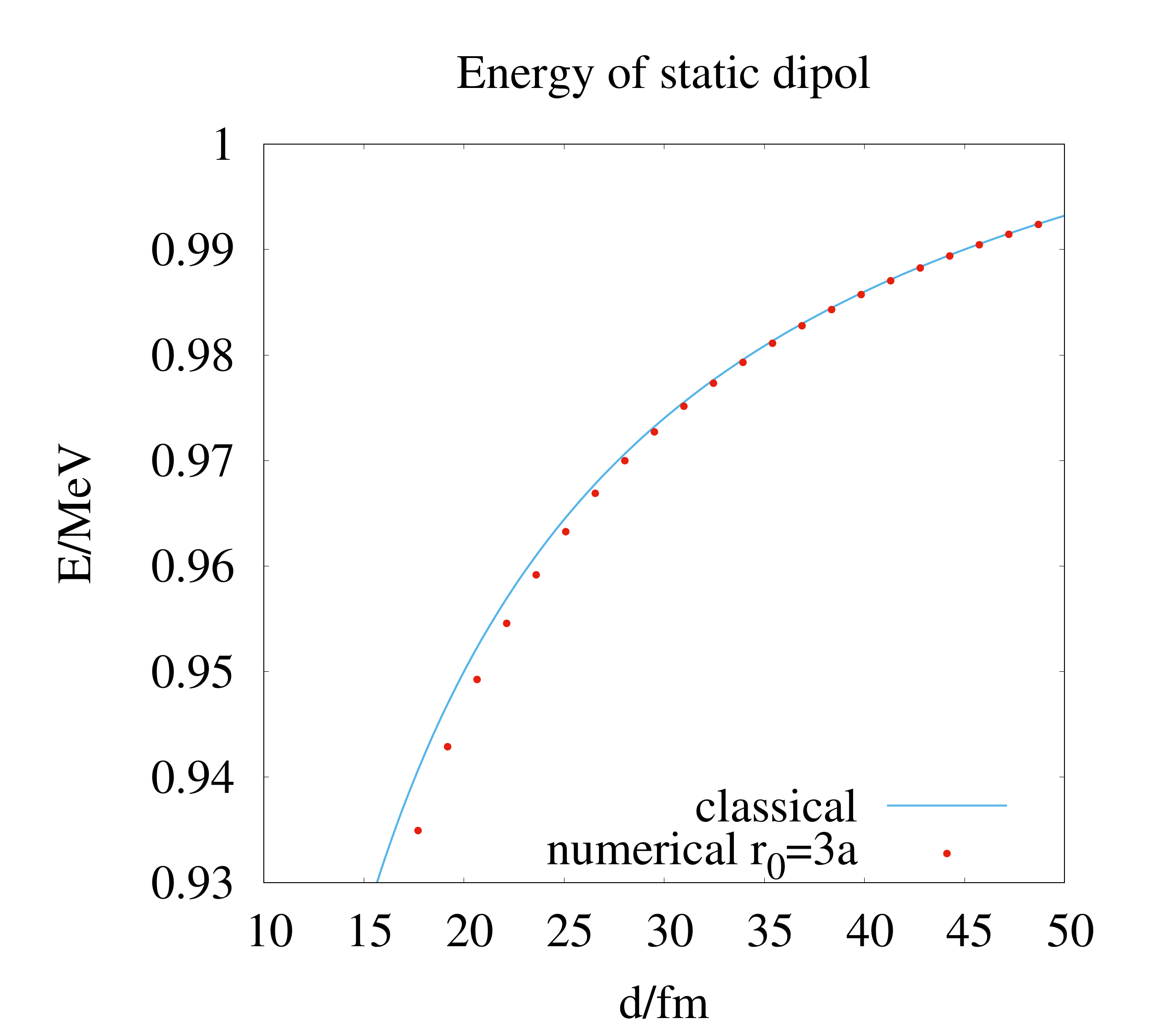}
\caption{Comparison between the classical Coulomb potential $E_\mathrm{cl}(d)$ and the numerical values of the energy $E(d)$ according to Eq.~(\ref{Enumerisch}) of a static electron-positron pair as a function of the separation $d$ between the two charges. Due to the finite size of solitons we observe deviations of the energy of the static dipole (red circles) from the classical value for point-like sources (full blue line).}
\label{potfig}
\end{figure}

From the analytical calculations in Ref.~\cite{Faber:2002nw} we know that our model respects the inhomogeneous Maxwell equations. Therefore, we expect for point-like charges the classical Coulombic behaviour at large distances $d$ between the charges of the dipole. This reads after adjusting the asymptotic energies
\begin{equation}\label{classEvond}
E_\mathrm{cl}(d)=E_\infty-\frac{\alpha_f\hbar c_0}{d}\quad\textrm{with}\quad\alpha_f=137.036, \hbar c_0=197.327\,\textrm{MeV fm}.
\end{equation}
In Fig.~\ref{potfig} we compare $E_\mathrm{cl}(d)$ to the numerical results for the energy $E(d)$, see Eq.~(\ref{Enumerisch}), of the dipole in the spin singlet state. The left diagram depicts the comparison in the region $24~a\le \bar d\le420~a$ in steps $\Delta \bar d=2\,a$. There are no data for distances smaller than $d=24~a$ for $a=r_0/3=0.7377\,$fm, since the soliton pair cannot be stabilised just by fixing the soliton field to $Q=1$ at the positions of soliton centers. The small deviations of the numerical values $E(d)$ from $E_\mathrm{cl}(d)$ at $d<50~$fm are magnified in the right diagram of Fig.~\ref{potfig}.

\begin{figure}[h]
\centering
\includegraphics[scale=0.3]{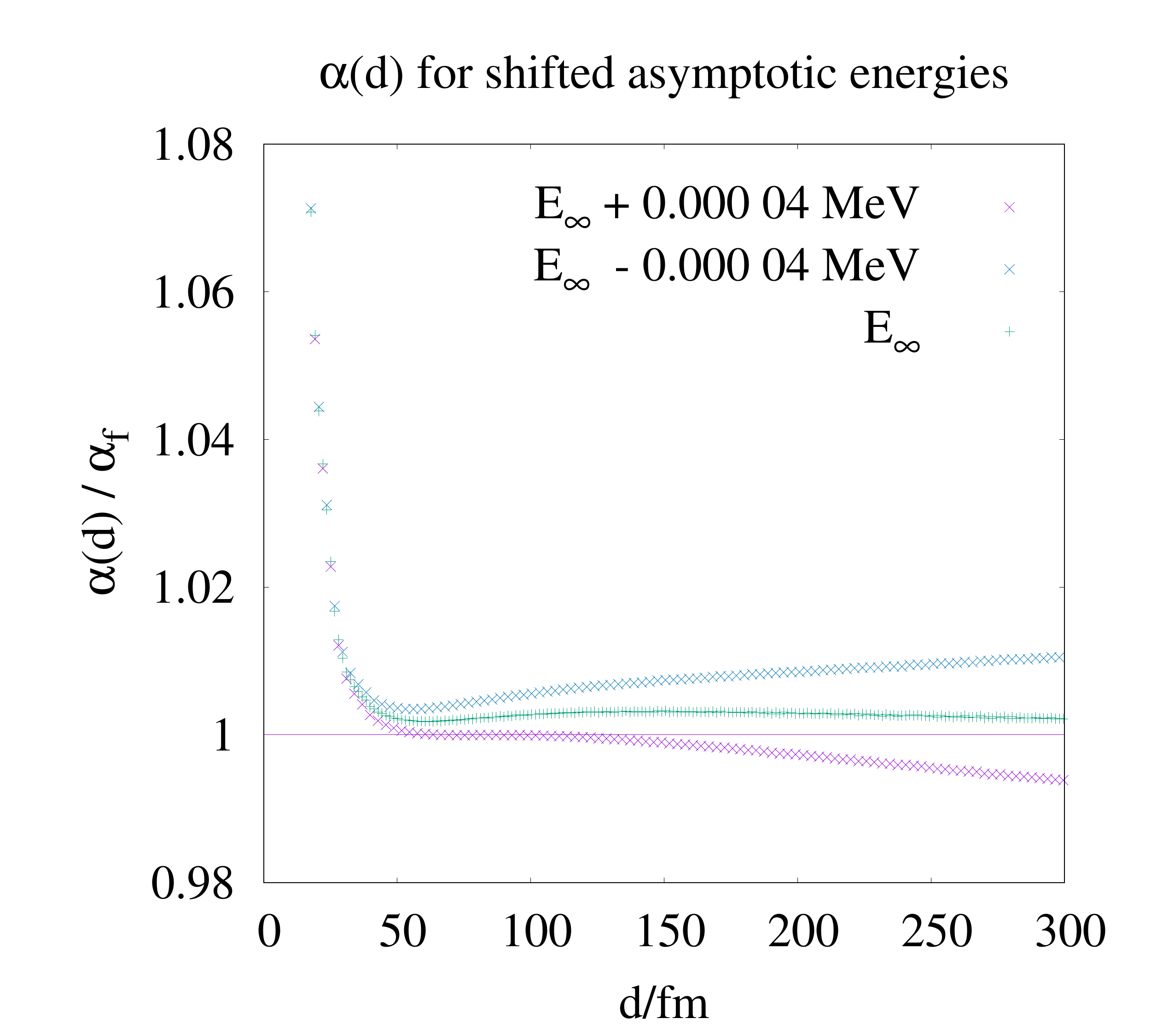}
\caption{Influence of the choice of $E_\infty$ to the asymptotic behaviour of $\alpha(d)$ according to Eq.~(\ref{Evond}). Observe that tiny variations of $E_\infty$ lead to wrong asymptotic behaviour at large charge separations.}
\label{Einftest}
\end{figure}

We describe these deviations by a $d$-dependence of the fine structure constant
\begin{equation}\label{Evond}
E(d)=E_\infty-\frac{\alpha(d)\,\hbar c_0}{d}.
\end{equation}
In Fig.~\ref{Einftest} we can nicely observe whether the asymptotic energy $E_\infty$ is well chosen, as tiny variations of $E_\infty$ destroy the asymptotic behaviour of $\alpha(d)$ at large $d$. The running of the coupling $\alpha(d)$ already detected in Fig.~\ref{potfig} is clearly seen in Fig.~\ref{Einftest} below distances of 50\,fm.

\begin{figure}[h]
\centering
\includegraphics[scale=0.3]{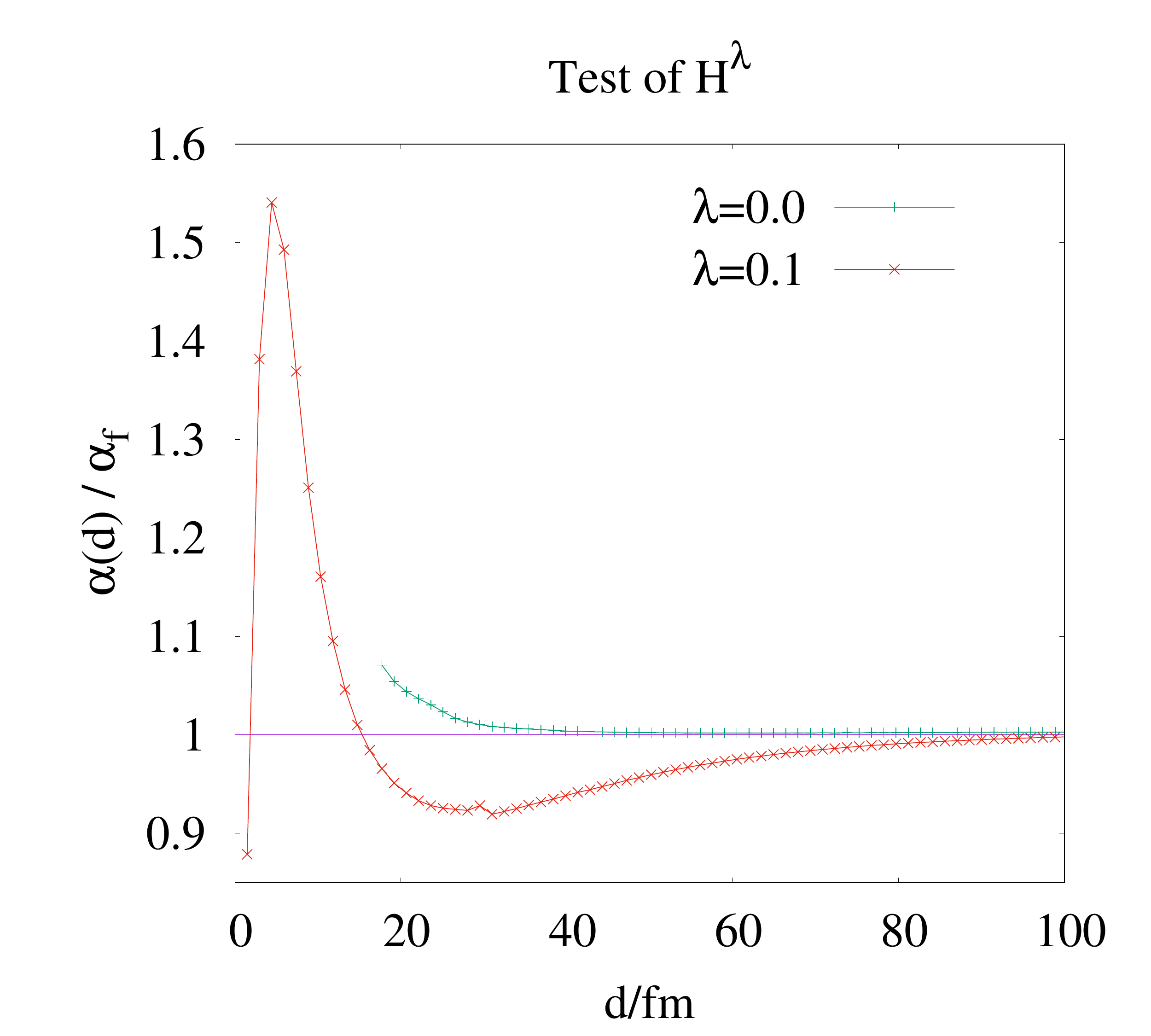}
\caption{The suppression of photonic excitations by the $\lambda$-dependent term in Eq.~(\ref{massTerm}) allows to approach shorter distances but leads to an unphysical reduction of $\alpha(d)$ around distances of 30~fm, as can be clearly observed for $\lambda=0.1/a^2$.}
\label{stiffalpha}
\end{figure}

To be able to reach smaller values of $d$ before the attraction gets too strong and the dipole is collapsing, Ref.~\cite{Anmasser2021} suggested to minimise the energy funcional $E_\mathrm{min}$ of Eq.~(\ref{massTerm}) which suppresses photonic excitations. In the computations, see Fig.~\ref{stiffalpha}, it turns out that it helps to approach shorter distances, but at distances of around 30\,fm it may lead to unphysical stiffness of the fields and to a strong reduction of $\alpha(d)$.
This unphysical behaviour starts to be observable around $\lambda=0.005/a^2$. Looking carefully at Fig.~\ref{Einftest} we see a similiar but very tiny additional stiffness also in the $\lambda=0$ results. It is not yet clear whether the minimum in $\alpha(d)/\alpha_f$ at $d\approx50$\~fm is a physical effect or an error of the approximation, e.g. of the boundary conditions. Due to its small size this effect is not really visible in Fig.~\ref{stiffalpha} at $\lambda=0$.

\begin{figure}[h!]
\centering
\includegraphics[scale=0.3]{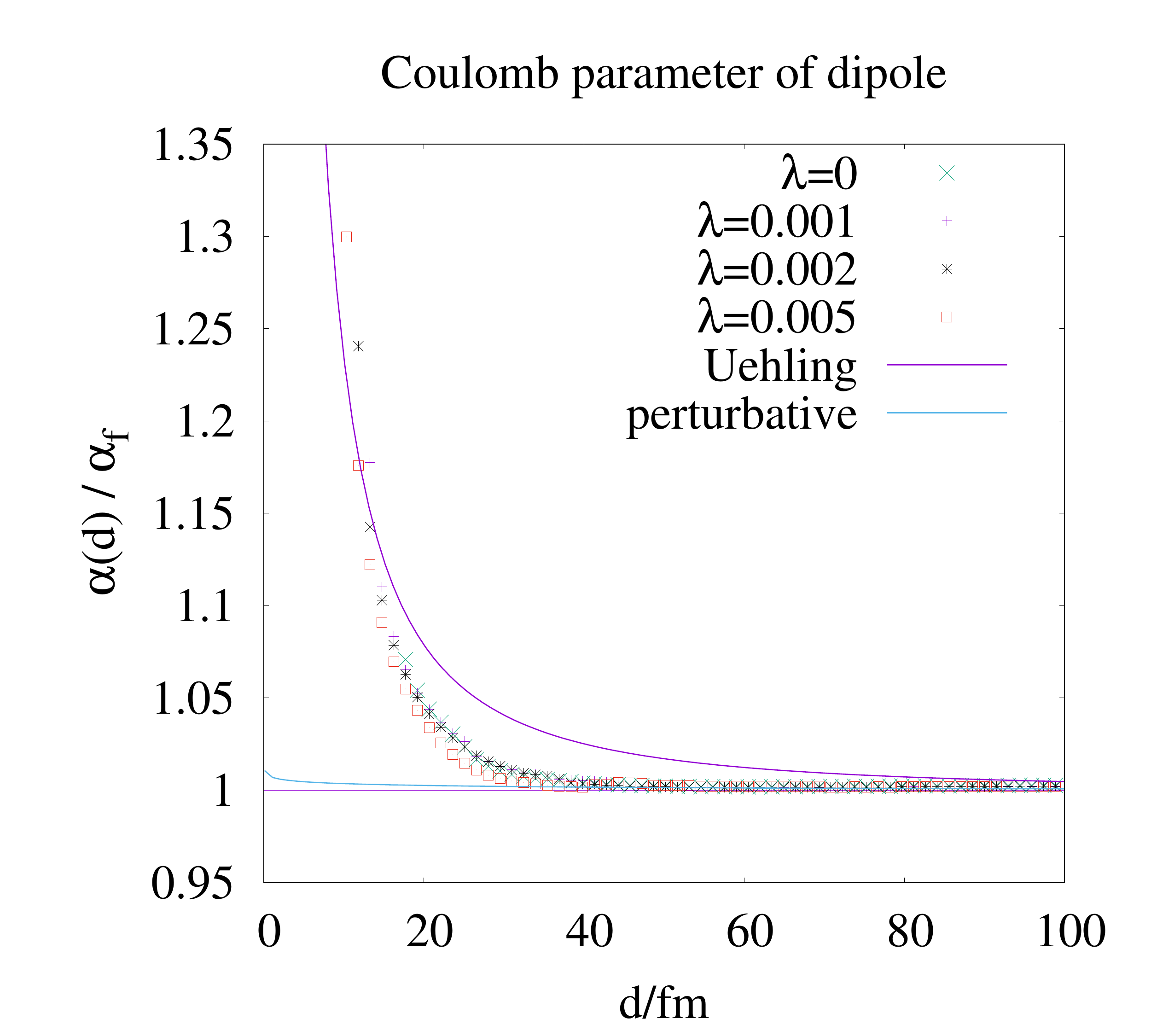}
\caption{The strong attraction at short distances does not allow to stabilize the distance $d$ of the charges just by fixing the centers of solitons with the energy functional~(\ref{Enumerisch}). One can get to shorter distances using the functional of Eq.~(\ref{massTerm}) for the minimisation procedure. For tiny values of $\lambda$ the potential is only weakly distorted. The rise of $\alpha(d)$ is compared with the result of first order perturbation theory~(\ref{pertalpha}) in QED and its long distance approximation, the Uehling potential~(\ref{PeskinAppr}).}
\label{mehrereLambdas}
\end{figure}

For tiny $\lambda\le 0.005/a^2$ we can get to smaller $d$-values. The corresponding data is plotted in Fig.~\ref{mehrereLambdas}. It is interesting to compare the rise of the coupling at small $d$ to the perturbative calculation in QED as described in ref.~\cite{peskin}. For comparison we show the prediction of first order perturbation theory and its long distance approximation which is well known as Uehling potential.

\begin{figure}[h!]
\centering
\includegraphics[scale=0.75]{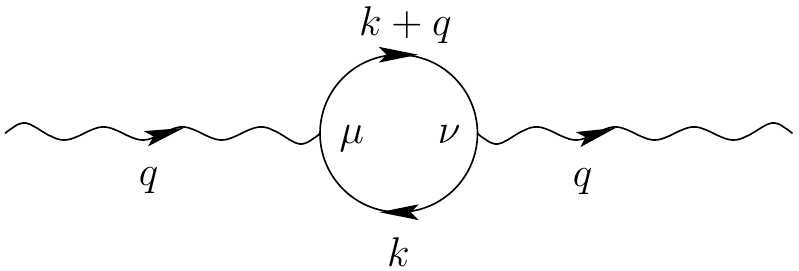}
\caption{Electron-loop diagram as contribution to the photon propagator at four-momentum $q$ in first order perturbation theory.}
\label{FotonSelbst1}
\end{figure}

In QED the interaction between charges is described by photon exchange. The algebraic form of the photon propagator determines therefore the strength of the interaction. The perturbative corrections to the tree level photon propagator are divergent. Already the first order contribution, shown by the electron-loop diagram in Fig.~\ref{FotonSelbst1}, is divergent and needs regularisation and renormalisation.
A perfect regularisation method is dimensional regularisation. Its success is understood from the conservation of symmetries besides scale invariance. In the renormalisation procedure an infinite photon self-energy is subtracted. It is adjusted in such a way that the divergence of the photon propagator is cancelled. Invariance under the finite terms which are subtracted in this procedure requests a modification of the electric charge at large momentum transfers and small distances. This is usually interpreted as a modification of the photon structure by virtual electron-positron pairs. This prediction of perturbation theory leads to a radial dependent fine structure constant
\begin{equation}\label{pertalpha}
\alpha(r)=\alpha_f\Big\{1+
\underbrace{\frac{2\alpha_f}{3\pi}
\int_{1}^\infty\mathrm d\gamma\frac{\mathrm e^{-2\mu\gamma r}}{\gamma}
\sqrt{1-\frac{1}{\gamma^2}}\Big(1+\frac{1}{2\gamma^2}\Big)}_{\Delta(r)}\Big\},
\end{equation}
where $\mu$ is the reduced Compton wavelength of electrons and $\gamma$ is the relativistic $\gamma$ factor for the mass of a moving particle. The correction can be evaluated by numerical integration and leads to the curve ``perturbative'' in Fig.~\ref{mehrereLambdas}. The approximation for large $\mu r$, the Uehling potential, reads~\cite{peskin}
\begin{equation}\label{PeskinAppr}
\alpha(r)=\alpha_f\Big\{1+\frac{\alpha_f}{\sqrt{2\pi}}
\frac{\mathrm e^{-2\mu r}}{(2\mu r)^{3/2}}\Big\}.
\end{equation}
 It is depicted for comparison in Fig.~\ref{mehrereLambdas}.

\section{Conclusion}\label{Sec:Conclusions}
In the model of topological particles we can describe monopoles and their interaction without any divergencies. Since analytic solutions of this model are known for one-particle systems only, we have to analyse interacting systems numerically. In this first calculation we investigate the classical monopole-antimonopole potential in the spin zero state. This is an axial symmetric configuration which can be studied in the rz-plane in cylindrical coordinates. At large separation $d$ of the dipole the potential is purely Coulombic. The stable soliton states have finite size $r_0$ of the order of the classical electron radius. This leads to a running of the charge at short distances. It is a natural question how this running compares to the running of the coupling in QED. The increase of the coupling appears at roughly the same length scale as predicted by perturbative QED.

In further calculations the accuracy of the calculation should be enlarged using adaptive lattices. This would allow to diminish the influence of boundary effects. Since these first numerical calculations are done in the static limit one should expect major modifications for more realistic dynamical scenarios. Especially interesting would be to insert the potential which results from such more detailed calculation in the Schrödinger and Dirac equation~\cite{Faber_2022}. This would result in shifts of the Bohr energies which could be compared with the Lamb shift. Further calculations should be done in the spin one state and for the repulsive system of two equal charges where the cylindrical symmetry is lost.

\bibliography{literatur}
\bibliographystyle{unsrt}

\appendix
\section{More details of perturbative calculation}
The free propagator of a photon with momentum $q_\mu$, see the Feynman rule~(\ref{PhotonProp}) is transversal, i.e. proportional to the transversal projection operator $^T\!P^{\mu\nu}:=g^{\mu\nu}-\frac{q^\mu q^\nu}{q^2}$. Due to the Ward identity it remains transversal to all orders of magnitude. Summing up the infinite geometrical serious of 1-particle-irreducible diagrams
\begin{equation}
  \mathrm i\Pi^{\mu\nu}=\mathrm iq^2\,\Pi(q^2)\,^T\!P^{\mu\nu}
\end{equation}
ends up in the expression
\begin{equation}\begin{aligned}\label{propKorr}
&\includegraphics[scale=0.4]{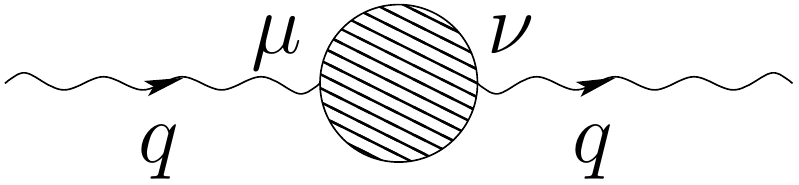}
=\mathrm iD^{\prime\mu\nu}(q)=\\
&\hspace{10mm}=\mathrm iD^{\mu\nu}(q)+\mathrm iD^{\mu\kappa}(q)
\,\mathrm i\Pi^{\kappa\lambda}(q)\,\mathrm iD^{\lambda\nu}(q)+\dots
=-\mathrm i\,\frac{^T\!P^{\mu\nu}}{q^2[1-\Pi(q^2)]}
\end{aligned}\end{equation}
for the photon propagator, where the self-energy $\Pi(q^2)$ can be determined in perturbation theory as a function of
\begin{equation}\label{4erImpulsQ}
q^2=q_0^2-\vec q^{\;2},
\end{equation}
see Sect.~\ref{Sec-PhotonProp} resulting in expression~(\ref{renormPi}) for $\Pi(q^2)$.

In QED the Coulomb interaction is described by one-photon exchange. This is possible, since the free photon propagator appears also in the momentum representation of the Coulomb potential
\begin{equation}\label{FTVr}
V(\vec q)\ist{VrErgeb}-\alpha_f\hbar c_0\,\frac{4\pi}{\vec q^{\;2}+\mu_\gamma^2}
\end{equation}
in the limit of vanishing photon mass
\begin{equation}\label{mugamma}
m_\gamma=:\frac{\mu_\gamma\hbar}{c_0}.
\end{equation}
From the comparison of the QED expression~(\ref{propKorr}) with the Fourier transform ~(\ref{FTVr}) of the Coulomb potential (\ref{VrErgeb}) we conclude that  the QED-modification of the Coulomb potential has to be determined in the region $q^2=-\vec q^{\;2}$, i.e.
\begin{equation}\label{region}
q_0=0\quad\Rightarrow\quad q^2\ist{4erImpulsQ}-\vec q^{\;2}=:-Q^2
\end{equation}
of the photon momentum in the off mass shell region. From Eq.~(\ref{propKorr}) we read the perturbative correction
\begin{equation}\label{CoulKorr}
-\frac{1}{q^2}\ist{region}\frac{1}{Q^2}\;\;\zu{propKorr}\;\;
\frac{1}{Q^2[1-\Pi(-Q^2)]}.
\end{equation}
Restricting the geometrical series which we have summed up in Eq.~(\ref{propKorr}) to the first term, we get the substitution
\begin{equation}\label{CoulKorr1O}
\frac{1}{Q^2}\;\;\zu{propKorr}\;\;\frac{1}{Q^2}[1+\Pi(-Q^2)].
\end{equation}

\begin{figure}[h!]
\centering
\includegraphics[scale=0.75]{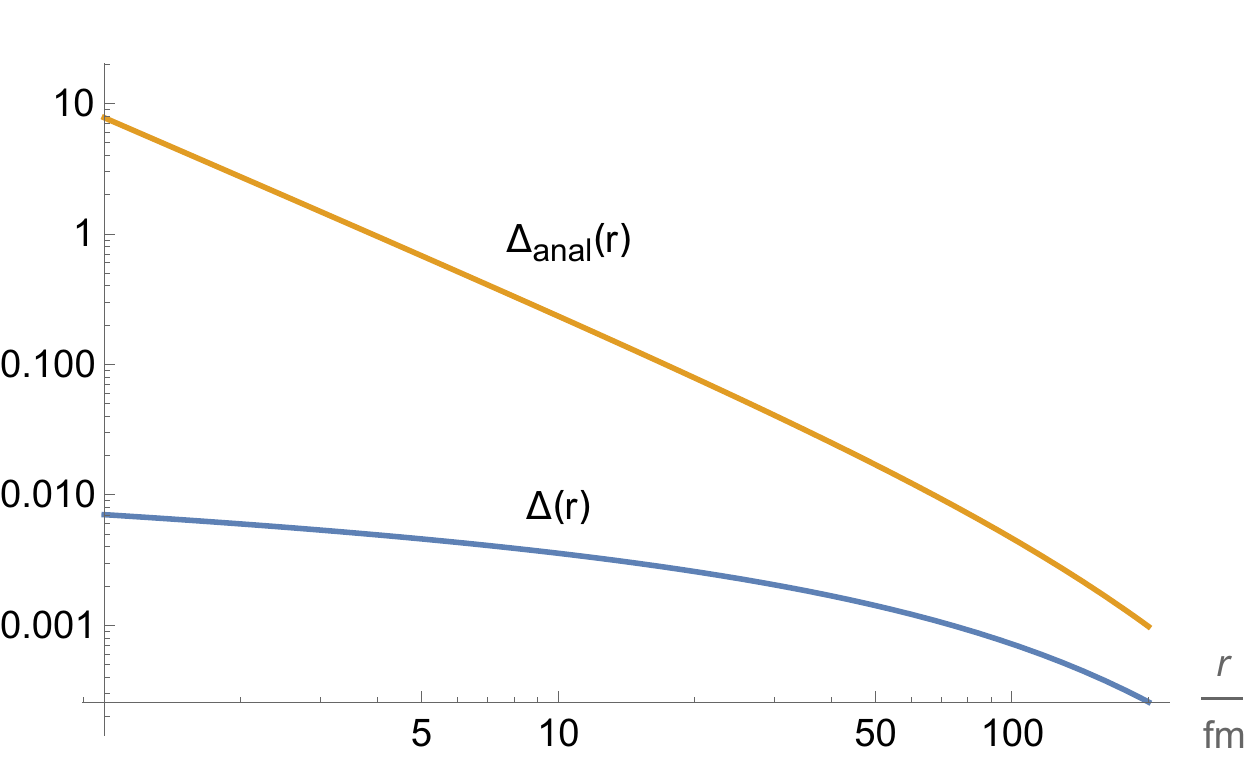}
\caption{Double logarithmic plot of the correction terms~(\ref{Korrektur}) and (\ref{PeskinAppr}) to $\alpha(r)$ of Eq.~(\ref{modalpha}) in the interesting $r$-region.}
\label{loglogCorr}
\end{figure}

If we insert this perturbative modification into the Coulomb potential
\begin{equation}\label{VrErgeb}
V(r):=-\frac{\alpha_f\hbar c_0}{r}\mathrm e^{-\mu_\gamma r}=
-\frac{\alpha_f\hbar c_0}{r}\frac{1}{\mathrm i\pi}
\int_{-\infty}^\infty\mathrm dQ
\frac{Q\,\mathrm e^{\mathrm iQr}}{Q^2+\mu_\gamma^2}
\end{equation}
we get the QED-correction to the potential at $\mu_\gamma=0$
\begin{equation}\label{DeltaV}
\delta V(r)\iist{VrErgeb}{CoulKorr1O}
-\frac{\alpha_f\hbar c_0}{r}\frac{1}{\mathrm i\pi}
\int_{-\infty}^\infty\mathrm dQ\frac{\mathrm e^{\mathrm iQr}}{Q}\Pi(-Q^2)\Big\}.
\end{equation}
First order perturbation theory results in the photon self-energy
\begin{equation}\label{Einschleifen}
\Pi(q^2)\ist{renormPi}\frac{2\alpha_f}{\pi}\int_0^1\mathrm dx\;
x(1-x)\log\Big[1-x(1-x)\frac{q^2}{\mu^2})\Big]
\end{equation}
and the correction term to the potential
\begin{equation}\label{DeltaVpert}
\delta V(r)\iist{DeltaV}{Einschleifen}
\mathrm i\,\frac{2\alpha_f^2\hbar c_0}{\pi^2r}
\int_{-\infty}^\infty\mathrm dQ\frac{\mathrm e^{\mathrm iQr}}{Q}
\int_0^1\mathrm dx\,x(1-x)\log\Big[1+x(1-x)\,\frac{Q^2}{\mu^2})\Big]\Big\}.
\end{equation}
Due to the antisymmetry of the integrand only the odd term, the sin-function in the exponential, should contribute to the $Q$-integral. But due to the strong oscillation of the sin-function the integral is difficult to evaluate. It is much easier to evaluate the $Q$-integral in the complex plane. The pole from the $1/Q$-term is suppressed by the vanishing logarithm at $Q=0$. But on the imaginary $Q$-axis there are branch cuts in the regions of $Q$-values with $|\textrm{Im }Q|\ge2\mu$. For these $Q$-values there is a region of $x(1-x)$-values, around the maximal value $1/4$, where the logarithm gets negative. Physically, this is the region of positive $q^2=-Q^2$, where annihilation takes place and the energy is concentrated in the mass of the fused object which cannot be lower than $2\mu$. For the evaluation of the integral in the complex plane we use the absence of poles. Therefore, the complex integral can be shifted as much as possible to the upper half-plane where the contributions of the two quarter circles in the first and the second quadrant are exponentially suppressed. The branch cut has to be circumvented by two straight paths at $\textrm{Re }Q=0$ from $\textrm{Im }Q:=q=\infty$ down to $q=2\mu$ and back to $+\infty$. These two straight paths give only an imaginary contribution since the two real contributions are equal and cancel due to the opposite integration direction. The argument
\begin{equation}\label{argLog}
z:\ist{DeltaVpert}=1+x(1-x)\,Q^2/\mu^2=1-x(1-x)\,q^2/\mu^2,\quad Q=\mathrm i q
\end{equation}
of the logarithm in Eq.~(\ref{DeltaVpert}) is negative in the interval
\begin{equation}\label{grenzenx}
x_{1,2}=\frac{1}{2}\Big(1\pm\sqrt{1-\frac{4\mu^2}{q^2}}\big)
:=\frac{1}{2}(1\pm\beta)\quad\textrm{with}\quad
\beta:=\sqrt{1-\frac{4\mu^2}{q^2}},
\end{equation}
which is characterised by real values of the velocity $\beta$ of a fusing particle pair. In this region of real $\beta$-values the imaginary part of
\begin{equation}\label{LogVerzw}
\log z=\log r\pm\mathrm i\pi\quad\textrm{for}\quad z=-r\pm\mathrm i0
\end{equation}
has opposite sign at the opposite sites $Q=\pm\eta+\mathrm i q$ of the branch cut
\begin{equation}\label{Logz}
\textrm{Im }\log z\ist{LogVerzw}\pi\,\textrm{sign(Im }z)
\ist{argLog}\pi\,\textrm{sign(Im }Q^2)
\ist{argLog}\pi\,\textrm{sign(Re }Q)
\end{equation}
depending on the infinitesimal real part of $Q$. The region with negative real part of $z$ contributes therefore in the first quadrant ($Q=+\eta+\mathrm i q$) with
\begin{align}\nonumber
\textrm{Im }&\int_0^1\mathrm dx\,x(1-x)\log\Big[1+x(1-x)\,\frac{Q^2}{\mu^2}\Big]
\ist{grenzenx}\pi\int_{\frac{1}{2}(1-\beta)}^{\frac{1}{2}(1+\beta)}\mathrm dx\;x(1-x)
=\\&\tist{u=x-1/2}\label{xIntegral}
\pi\underbrace{\int_{-\beta/2}^{\beta/2}\mathrm du\Big(\frac{1}{4}-u^2\Big)}
_{\beta/4-\beta^3/12}=\pi\,\frac{\beta}{12}(3-\beta^2)=\\\nonumber
&\ist{grenzenx}\frac{\pi}{6}\,
\sqrt{1-\frac{4\mu^2}{q^2}}\Big(1+\frac{2\mu^2}{q^2}\Big).
\end{align}
Since the straight path in the second quadrant contributes with the opposite sign we have to insert the double of the result~(\ref{xIntegral}) in $\delta V(r)$
\begin{equation}\label{DeltaVres}
\delta V(r)\iist{DeltaVpert}{xIntegral}
-\frac{\alpha_f\hbar c_0}{r}\frac{2\alpha_f}{3\pi}
\int_{2\mu}^\infty\mathrm dq\frac{\mathrm e^{-qr}}{q}
\sqrt{1-\frac{4\mu^2}{q^2}}\Big(1+\frac{2\mu^2}{q^2}\Big).
\end{equation}
There are two scales in the $q$-integral, the mass scale $2\mu$ and the distance $r$. It is convenient to go to the dimensionless relativistic momentum scale $\gamma=q/(2\mu)$
\begin{equation}\label{DeltaVgamma}
\delta V(r)\ist{DeltaVres}
-\frac{\alpha_f\hbar c_0}{r}\frac{2\alpha_f}{3\pi}
\int_{1}^\infty\mathrm d\gamma\frac{\mathrm e^{-2\mu\gamma r}}{\gamma}
\sqrt{1-\frac{1}{\gamma^2}}\Big(1+\frac{1}{2\gamma^2}\Big).
\end{equation}
The radial dependent fine structure constant reads therefore
\begin{equation}\label{modalpha}
\alpha(r)=\ist{DeltaVres}\alpha_f\Big\{1+
\underbrace{\frac{2\alpha_f}{3\pi}
\int_{1}^\infty\mathrm d\gamma\frac{\mathrm e^{-2\mu\gamma r}}{\gamma}
\sqrt{1-\frac{1}{\gamma^2}}\Big(1+\frac{1}{2\gamma^2}\Big)}_{\Delta(r)}\Big\}.
\end{equation}
The prefactor $\frac{2\alpha_f}{3\pi}$=0.00154 of the correction term
\begin{equation}\label{Korrektur}
\Delta(r):=\frac{2\alpha_f}{3\pi}
\int_{1}^\infty\mathrm d\gamma\frac{\mathrm e^{-2\mu\gamma r}}{\gamma}
\sqrt{1-\frac{1}{\gamma^2}}\Big(1+\frac{1}{2\gamma^2}\Big)
\end{equation}
is rather small. From the exponential factor $\mathrm e^{-2\mu\gamma r}$ at $\gamma=1$ we expect therefore sufficiently large correction terms $\Delta(r)$ for $r$-values much smaller than the reduced Compton wave length $\frac{1}{2\mu}=193\,$fm of an electron pair, i.e. $\frac{r}{2\mu}\ll1$. The function $\frac{\beta}{\gamma}=\frac{1}{\gamma}\sqrt{1-\frac{1}{\gamma^2}}$ in the argument of the $\gamma$-integral is dominated by $\beta$ at $\gamma=1$ and slowly decreasing with $1/\gamma$ for large $\gamma$ with a maximum of 0.5 at $\gamma=\sqrt{2}$. The factor $1+\frac{1}{2\gamma^2}$ leads to a small increase of the maximum and a small shift to lower $\gamma$ only. The numerical evaluation of the correction term~(\ref{Korrektur}) is depicted in Fig.~\ref{loglogCorr} together with an analytical approximation, valid for $r\gg\frac{1}{2\mu}$ only.

To get an approximation for large $2\mu r$ we can substitute $\gamma=t+1$ in Eq.~(\ref{Korrektur})
\begin{equation}\label{KorrekturGrosseR}
\Delta(r)\ist{Korrektur}\frac{2\alpha_f}{3\pi}\mathrm e^{-2\mu r}
\int_{0}^\infty\mathrm dt\,\frac{\mathrm e^{-2\mu rt}}{t+1}
\sqrt{1-\frac{1}{(t+1)^2}}\Big(1+\frac{1}{2(t+1)^2}\Big).
\end{equation}
With the expansion
\begin{equation}\label{tEntw}
\frac{1}{t+1}\sqrt{1-\frac{1}{(t+1)^2}}\Big(1+\frac{1}{2(t+1)^2}\Big)\simeq
3\sqrt{\frac{t}{2}}-\frac{29}{2}\left(\frac{t}{2}\right)^{3/2}
\end{equation}
we arrive at the integrals
\begin{equation}\label{KorrekturApprox}
\Delta(r)\simeq\frac{\sqrt 2\alpha_f}{\pi}\mathrm e^{-2\mu r}
\int_{0}^\infty\mathrm dt\,\mathrm e^{-2\mu rt}\sqrt{t}
=\frac{\sqrt 2\alpha_f}{\pi}\frac{\mathrm e^{-2\mu r}}{(2\mu r)^{3/2}}
\underbrace{\int_{0}^\infty\mathrm dx\,\mathrm e^{-x}\sqrt{x}}_{\sqrt{\pi}/2}
\end{equation}
and the analytical approximation
\begin{equation}\label{PeskinAppr}
\Delta(r)\simeq\Delta_\mathrm{anal}(r):=\frac{\alpha_f}{\sqrt{2\pi}}
\frac{\mathrm e^{-2\mu r}}{(2\mu r)^{3/2}}
\end{equation}
given in~\cite{peskin}, valid for $2\mu r\gg1$ only. At $2\mu r=1$ we get $\Delta_\mathrm{anal}(\frac{1}{2\mu})=0.00107$ which is by a factor of 4 too large compared to the numerical value $\Delta(\frac{1}{2\mu})=0.00028$. A comparison of the two corrections $\Delta(r)$ and $\Delta_\mathrm{anal}(r)$ is shown in a double logarithmic plot in Fig.~\ref{loglogCorr}.

\section{Perturbative correction}\label{Sec-PhotonProp}
The processes, we want to describe, are electron-electron, positron-positron, electron-positron scattering and electron-positron annihilation. They differ in the particle energies and transferred momenta.

According to the Feynman rules\\
\begin{minipage}[c]{37mm}
\includegraphics[scale=1.0]{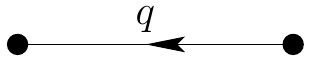}\quad
\end{minipage}
\begin{minipage}[c]{103mm}
\begin{flalign}
&\mathrm i\,S(q)=\mathrm i\frac{\slashed q+\mu}{q^2-\mu^2+\mathrm i\varepsilon}&
\end{flalign}
\end{minipage}\\
\begin{minipage}[c]{37mm}
\includegraphics[scale=1.0]{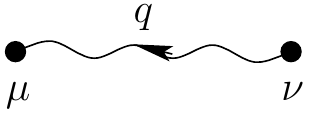}\quad
\end{minipage}
\begin{minipage}[c]{103mm}
\begin{flalign}\label{PhotonProp}
&\mathrm i\,D^{\mu\nu}(q)=-\mathrm i
\frac{1}{q^2+\mathrm i\varepsilon}[g^{\mu\nu}-\frac{q^\mu q^\nu}{q^2}]&
\end{flalign}
\end{minipage}\\
\begin{minipage}[c]{40mm}
\includegraphics[scale=1.0,bb=0 25 90 90]{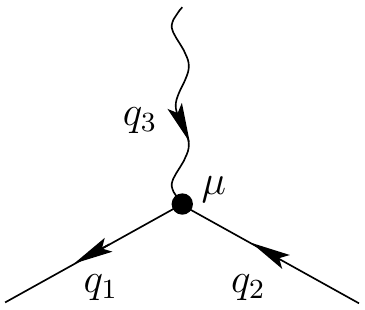}
\end{minipage}
\begin{minipage}[c]{100mm}
\begin{flalign}\label{QGVert}
&-\mathrm ie_0\gamma_\mu\quad\textrm{with}\quad\sum_iq_i=0,&
\end{flalign}
\end{minipage}\\[20mm]
the rules for the loop integrals $\int\frac{\mathrm d^4k}{(2\pi)^4}$ and the factor $(-1)$ for fermion loops, we get the contribution of the electron-loop of Fig.~\ref{FotonSelbst1}
\begin{equation}\label{1ElektronSchleife}
\mathrm i\Pi_2^{\mu\nu}(q):=-(-\mathrm ie_0)^2\int\frac{\mathrm d^4k}{(2\pi)^4}
\mathrm{tr}\Big[\gamma^\mu\frac{\mathrm i}{\slashed k-\mu}\gamma^\nu
\frac{\mathrm i}{\slashed k+\slashed q-\mu}\Big].
\end{equation}
to the photon propagator.

The regularisation of this divergent contribution to the photon propagator in one-loop approximation is convenient in Euclidean space and $\epsilon=4-D$ dimensions. The factorials in the dimension $D$ are interpolated by the Gamma function with poles at $0$ and all negative integers. The factors depending on $\epsilon$ and their expansion read
\begin{equation}\label{dimFakt}
(4\pi)^{-D/2}\Gamma(\frac{\epsilon}{2})=\frac{1}{(4\pi)^2}\left[
\frac{2}{\epsilon}-\gamma_E+\ln4\pi+\mathcal O(\epsilon)\right]
\end{equation}
where the Euler-Mascheroni constant $\gamma_E$ appears in $\Gamma(\frac{\epsilon}{2})=\frac{2}{\epsilon}-\gamma_E+\mathcal O(\epsilon)$.

After a long calculation, the expression for the photon self-energy $\Pi(q^2)$ reads in one-loop approximation~\cite{peskin}~\footnote{ corresponding to second order in the electric charge}
\begin{equation}\label{zweiteOrd}
\Pi_2(q^2)=-\frac{2\alpha_f}{\pi}\int_0^1\mathrm dx\;x(1-x)
\Big[\frac{2}{\epsilon}-\gamma_E+\log(4\pi)
-\log(\frac{\mu^2}{[\mu]^2}-x(1-x)\frac{q^2}{[\mu]^2})\Big].
\end{equation}
with a loop of an electron of mass $m$, inverse Compton wave length $\mu=mc/\hbar$ and scale $[\mu]$. The behaviour of $\Pi(q^2)$ for $\epsilon\to0$ demonstrates the well-known logarithmic infinity of the bare electric charge. It can be removed by a subtraction of this infinity and can be accompanied by a subtraction of arbitrary finite terms. There is no rule in the theory what finite contribution should be subtracted. These finite terms can only be fixed by a comparison with experiments. Since we know from experiments the charge of electrons at rest, it is wise to adjust the finite constant to this observation and define the electron self-energy by 
\begin{equation}\begin{aligned}\label{renormPi}
\Pi(q^2):&=\Pi_2(q^2)-\Pi_2(0)\ist{zweiteOrd}\\
&=\frac{2\alpha_f}{\pi}\int_0^1\mathrm dx\;
x(1-x)\log\Big[1-x(1-x)\frac{q^2}{\mu^2})\Big].
\end{aligned}\end{equation}

This expression reveals the famous dependence of the electron self-energy from the transferred momentum, the running of the coupling. The sign of $q^2$ depends on the experiment and the time structure of the corresponding Feynman diagram, whether the virtual photon line is time-like or space-like, whether mainly energy or momentum are transferred.

\begin{figure}[h!]
\centering
\includegraphics[scale=0.9]{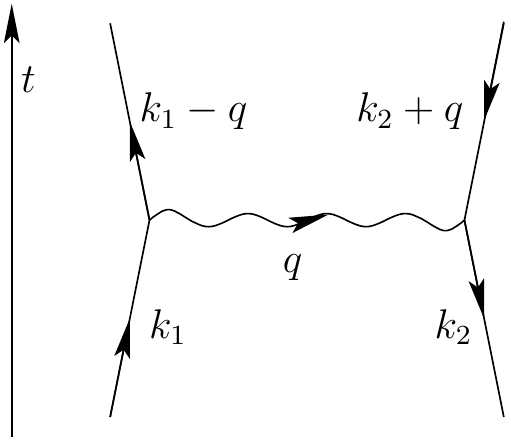}
\caption{Electron-positron scattering.}
\label{streuung}
\end{figure}

In a scattering experiment, see Fig.~\ref{streuung}, the transferred momentum is space-like
\begin{equation}\label{Qquadrat}
q^2=-Q^2.
\end{equation}

\end{document}